\documentclass[journal, twocolumn,twoside]{IEEEtran}

\usepackage{amssymb}
\usepackage{color}
\usepackage{cite}
\usepackage{graphicx,epsfig}
\usepackage[cmex10]{amsmath}
\usepackage{subfigure}
\usepackage{hyperref}
\usepackage{tablefootnote}
\usepackage{etoolbox}
\usepackage{color,soul}
\makeatletter
\patchcmd{\@makecaption}
{\scshape}
{}
{}
{}
\makeatletter
\patchcmd{\@makecaption}
{\\}
{.\ }
{}
{}
\makeatother

\newcommand{\rhoDL}{\rho^\text{DL}_{k_\text{d}}}
\newcommand{\rhoUL}{\rho^\text{UL}_{k_\text{u}}}
\newcommand{\rhoULell}{\rho^\text{UL}_{\ell}}
\newcommand{\rhoUU}{\rho^\text{UU}_{k_\text{d},k_\text{u}}}
\newcommand{\betaDL}{\beta^\text{DL}_{k_\text{d}}}
\newcommand{\betaUL}{\beta^\text{UL}_{k_\text{u}}}

\newcommand{\betaUU}{\beta^\text{UU}_{k_\text{d},k_\text{u}}}
\newcommand{\kd}{k_\text{d}}
\newcommand{\ku}{k_\text{u}}

\newtheorem{theorem}{Theorem}
\newtheorem{theorem2}{Theorem2}
\newtheorem{lemma}[theorem]{Lemma}
\newtheorem{remark}[theorem2]{Remark}

\begin{document}
	\bstctlcite{IEEEexample:BSTcontrol}

	\title{Performance Analysis of Self-Interference Cancellation in Full-Duplex Massive MIMO Systems: Subtraction versus Spatial Suppression}

	\author{Soo-Min~Kim,~\IEEEmembership{Graduate~Student~Member,~IEEE,}
	Yeon-Geun~Lim,~\IEEEmembership{Member,~IEEE,}\\
	Linglong Dai,~\IEEEmembership{Fellow,~IEEE,}
	and Chan-Byoung~Chae,~\IEEEmembership{Fellow,~IEEE}

		\thanks{
			This research was supported by IITP grant funded by the Korea government (2018-0-00923, 2021-0-00486). An earlier version of this paper was presented, in part, at the IEEE Global Telecommunications Conference (GLOBECOM) in 2016~\cite{Lim_Globecom17}.}
		\thanks{S.-M. Kim and C.-B. Chae are with Yonsei University, Korea. Email: \{sm.kim, cbchae\}@yonsei.ac.kr. Y.-G. Lim is with Samsung Electronics, Korea. Email: yg.lim@samsung.com. L. Dai is with Tsinghua University, Beijing 100084, China. Email: daill@tsinghua.edu.cn. S.-M. Kim and Y.-G. Lim are co-1st authors.}
}
	
	\maketitle

	\begin{abstract}
		Massive multiple-input multiple-output (MIMO) and full-duplex (FD) are promising candidates for achieving the spectral efficiency to meet the needs of 5G communications. One essential key to realizing practical FD massive MIMO systems is how to effectively mitigate the self-interference (SI). Conventionally, however, the performance comparison of different SI methods by reflecting the actual channel characteristics was insufficient in the literature. Accordingly, this paper presents a performance analysis of SI cancellation (SIC) methods in FD massive MIMO systems. Analytical and numerical results confirm that, in an imperfect channel-estimation case, the ergodic rates performance of the spatial suppression in the uplink outperforms those of the SI subtraction, due to the correlation between the precoder and the estimation error of the SI channel. In addition, we discuss which method performs better under different given system constraints such as uplink and downlink sum rates, the total transmit power, and the power scaling law.
	\end{abstract}
	

	\IEEEpeerreviewmaketitle
	
	\section{Introduction}
	
	The fifth-generation (5G) communication systems can support various types of services, including enhanced mobile broadband (eMBB), massive machine-type communication, and ultra-reliable low-latency communication \cite{5G,Lim_WCM18}. As mobile data traffic is expected to increase, eMBB has become a crucial part of 5G communications. To support the tremendous data traffic of 5G, researchers should jointly consider key issues such as spectral efficiency (bps/Hz) and spectrum extension (Hz/cell). As researchers developed key issues over the past few decades, massive multiple-input multiple-output (MIMO) technology has been suggested as a candidate to achieve high spectral efficiency. At the same time, researchers~\cite{Sachin_fd,Kolo_fd} intending to double the spectrum extension without shifting the system operating frequency band have studied full-duplex (FD) transmission. 
	
	To maximize spectral efficiency up to tens of bps/Hz and conserve energy, the researchers in \cite{Marzetta_WC10,Yang, Rusek_SPMag_12, Ngo_TCOM12, Lim_TWC} have proposed massive MIMO with tens of antennas as a core technology of 5G. They considered using simple linear precoders and filters to mitigate the interferences in massive MIMO systems. One key challenge in commercializing massive MIMO systems is the estimation of the high-dimensional MIMO channel matrix. This paper focuses on the channel estimation based on practical channel characteristics \cite{Yang, Rusek_SPMag_12, Ngo_TCOM12}. Specifically, the authors in \cite{Yang, Rusek_SPMag_12} studied the downlink performance of the maximum ratio transmission (MRT) and zero-forcing (ZF) precoders while assuming either perfect or imperfect channel estimation at the base station (BS). The authors in \cite{Ngo_TCOM12} investigated the uplink performance of the maximum ratio combining (MRC), ZF, and minimum mean square error (MMSE) filters with perfect or imperfect channel estimation. 
	
	Another relevant technology is FD transmission, which, compared to half-duplex transmission, can double the spectrum efficiency~\cite{Kolo_fd, Chae_WCM21, Chae_access} and reduce transmission latency~\cite{Kim_PIEEE19}. In contrast to half-duplex transmission, self-interference (SI), which is a transmitted downlink signal but is directly received at the BS when receiving an uplink signal, should be mitigated. The level of SI is usually higher than that of the receiver noise floor, and thus, any FD system requires SI cancellation (SIC) in both the analog and digital domains~\cite{Suk_FD}. There are typically two types of classical SIC methods in a digital-domain, that is, SI subtraction and spatial suppression. Specifically, SI subtraction subtracts replicated interference signal from the input signal, while spatial suppression applies multi-antenna techniques that utilize extra degrees of freedom offered by spatial dimensions for linear receive and transmit filtering (It is now considered in the form of null-space projection (NSP)~\cite{NSP_10}).
	From the digital-domain SIC perspective, the authors in \cite{FDSI_TSP11} compared SI subtraction and spatial suppression in a MIMO relay system with a small number of antennas (up to four antennas) at the BS and user. They showed that the performance of the spatial suppression was depended on the rank of the SI channel. 
	The authors paid little attention to the performance analysis of practical massive MIMO systems, such as user rate and uplink/downlink performance, because they focused on introducing advantages and disadvantages for each scheme according to the number of antennas and the rank of the channel. In a practical small MIMO system, SI subtraction is a widely considered SIC technique owing to the limitation of the requirement of additional antennas. The researchers in \cite{MK_ACCESS,SM_ACCESS20,Kim_Dyspan18,park_ICC} prototyped SI subtraction on a software-defined radio platform in real time. They combined a dual-polarization antenna-based analog cancellation with SI subtraction and realized a spectral efficiency 1.9 times higher than that of the half-duplex transmission. Nevertheless, spatial suppression has received limited attention in small MIMO systems.
	
	As aforementioned, in 5G and beyond 5G systems, integrated techniques are required to efficiently fulfill the requirements of eMBB. 
	In particular, we should integrate spectrum extension technologies with massive MIMO to enhance both the data rate of the network and each user. Conventional massive MIMO systems assume a small number of antennas at the users (for example, a single antenna) with a per-user rate constraint, which is limited by the used modulation technologies and allocated bandwidth. 
	In addition, massive MIMO systems are typically launched in the sub-6-GHz range, whereas hybrid beamforming technologies are considered in the millimeter-wave range~\cite{ComMag_Roh14}. Therefore, FD is a potential solution to the spectrum extension of massive MIMO systems. In both~\cite{Bai_TWC17} and \cite{Koh_TWC18}, the researchers investigated asymptotic ergodic sum rates and channel estimation methods in FD-assisted massive MIMO systems. They considered the condition of specific system constraints wherein the performance of an FD system outperforms that of a half-duplex system.
	The authors in~\cite{Yin_asi13} proposed spatial suppression-based algorithms such as extended ZF and extended MMSE precoders. As the conventional massive MIMO system serves a relatively smaller number of users than the number of antennas at the BS, the extended precoders can effectively mitigate SI by utilizing the dimensional feasibility in the spatial domain. As an interference-free channel was assumed in~\cite{Yin_asi13}, the analysis is not general. The researchers in~\cite{Min_TVT16} analyzed the ratio of transmit antennas and receive antennas in terms of the numbers of both the downlink and uplink users. The authors in \cite{Shojaeifard_TCOM17} utilized a spatial suppression precoder based on singular-value decomposition and proposed an uplink power control mechanism for cellular systems using a Poisson point process model. 
	Furthermore, the authors in~\cite{Huang1} showed the performance trade-off between the degree of freedom and the SI mitigation level, while the authors in~\cite{Huang2} presented a combination of spatial suppression and an analog domain cancellation solution that can relax the requirement of the number of antennas and reduce the complexity of SI cancellation. They demonstrated the performance analysis and feasibility of the spatial suppression method under imperfect channel estimation. However, there is insufficient explanation of whether SI suppression is optimal in the considered environments.
	
	In this paper, motivated by the potential of utilizing spatial suppression in massive MIMO systems but the lack of prior studies on its advantages compared to SI subtraction, we compared these two SIC methods in the digital domain. Prior works typically model SI channel coefficients as simple independent identically distributed (i.i.d.) complex Gaussian random variables. In this prior channel model, the researchers assumed that the uplink performances of both SIC methods were the same~\cite{Yin_asi13,Min_TVT16}. Thus, SI subtraction was always a better choice, because spatial suppression requires additional spatial resources at the downlink, which causes a difference in the number of antennas required to obtain a diversity gain for the downlink. In a practical channel, however, an SI channel would be highly correlated because of the higher power in the direct path. Therefore, in the imperfect channel, spatial suppression can be better for the uplink, depending on the magnitude of the channel estimation error. From this point of view, we analyzed the performance according to the diversity gain, number of antennas, and channel estimation error by performing the perfect channel analysis together. Motivated by~\cite{FDSI_TSP11}, in small FD MIMO systems, SI subtraction is better because the diversity gain for the downlink is limited because there are fewer antennas. However, in FD massive MIMO systems, because this limit is weaker, spatial suppression may be better. We mathematically prove this content accurately. In addition, the considered small FD MIMO system is not sufficient to meet the requirements of massive MIMO systems in beyond 5G. This paper presents a comparison of the two SIC methods with the following considerations in a system: i) channel estimation error, ii) spatial correlation, and iii) asymmetric data traffic with a non-full buffer, and power scaling law at the downlink. Based on the analysis and numerical results, we determined which method is better in a certain scenario. In our prior work~\cite{Lim_Globecom17}, two methods were compared in perfect and imperfect SI channel conditions, particularly in uncorrelated conditions. Furthermore, the two methods were compared through mathematical analysis and proofs in correlated conditions. The performance was also evaluated by considering the practical system constraints mentioned above. Through these analyses, results were obtained using a system-level simulator based on a 3D ray-tracing tool to obtain the real SI channel. The additive quantization noise model (AQNM) was also considered as a channel estimation error model to design the system more practically. Specifically, the main contributions of this study are summarized as follows.

	\begin{itemize}
		\item {\bf Performance analysis with imperfect channel estimation:} First, we provide the performance analysis to show that the uplink performance of spatial suppression is better than that of SI subtraction with imperfect channel estimation. The channel estimation error would be high, particularly in practical full-digital MIMO systems that usually use analog-to-digital converters (ADCs) with few bits (for example, 4-5 bits) for a cost-effective design~\cite{Bai_ISWCS13, Orhan_ITA15, Fan_COML15}. The SIC algorithms having improved performance with imperfect channel estimation are yet to be determined. We also derive the achievable ergodic sum rates of the two SIC methods, then we show that the uplink rate of spatial suppression is better than that of SI subtraction with an imperfect channel estimation error. 
		
		\item {\bf Performance analysis with spatial correlation:} As noted above, SI channels would be highly correlated. We prove this baseline channel condition by measuring the channel parameters with a 3D ray-tracing tool. Next, we investigate the impact of spatial correlation on the expected SI power. Based on this analysis, we conclude that the uplink rate of spatial suppression improves beyond that of SI subtraction as the spatial correlation increases. Notably, in the conventional analysis, these correlation terms of the SI channel are assumed as independent conditions~\cite{Yin_asi13,Min_TVT16}. 
		
		\item {\bf Performance comparison with system constraints:} Based on our analysis, we compare the two SIC methods when the system constraints are considered. These constraints include the ratio of downlink and uplink traffic, ratio of transmit and receive RF chains, transmit power at the BS, and power scaling law at the BS. Because of the performance trade-off between these two SIC schemes in a certain link direction, our analysis shows that each method exhibits different performances for different system constraints. These comparisons provide insights into the SIC selection criterion for certain system requirements.
	\end{itemize}

	The remainder of this paper is organized as follows. In Section~\ref{sec:system}, we introduce the system model and problem statement with respect to precoding/receive combining and SIC methods. In Section~\ref{sec:channel}, we present certain mathematical motivations and preliminaries that are useful for analysis. Then we compare the SIC algorithms under different channel conditions.
	In Section~\ref{sec:constraints}, we compare SIC algorithms with respect to the system constraints, such as the transmit power and traffic load. The numerical results are presented in Section~\ref{Section:C3_Section5}. Finally, Section~\ref{sec:conclusion} presents our conclusions.

	
			\begin{figure*}[t]
		\centering{\includegraphics[width=0.85\textwidth,keepaspectratio]{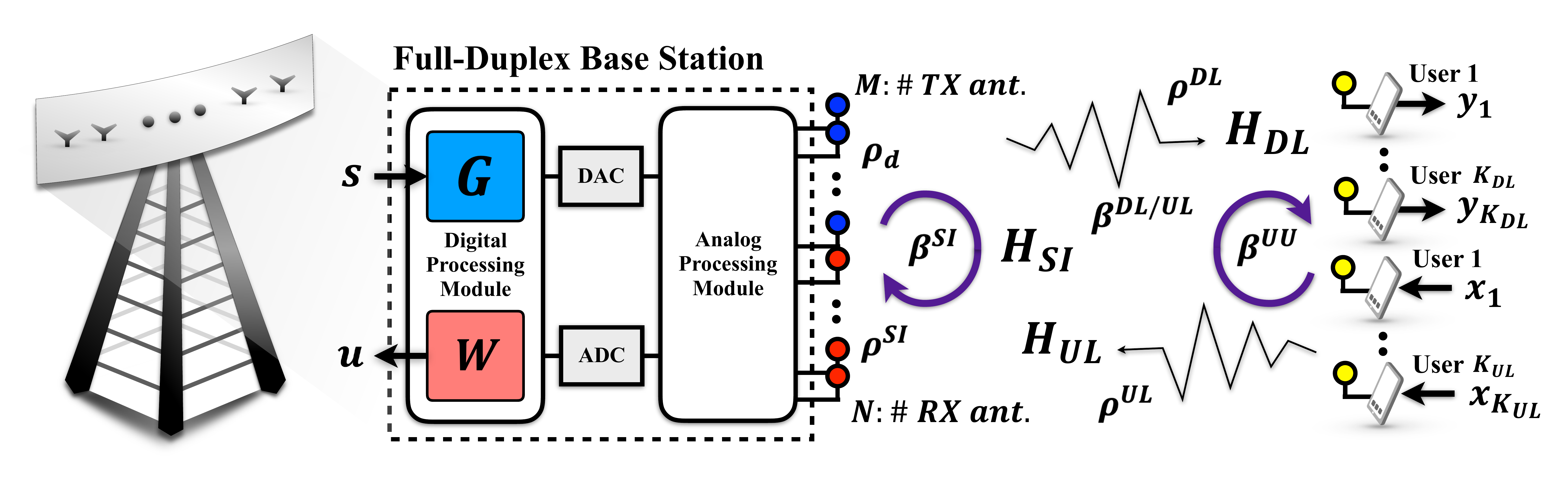}}
		\caption{A block diagram of an FD multi-user massive MIMO system.}
		\label{Sysmodel}
	\end{figure*}

	\section{System Model}
		\label{sec:system}

	
	In this section, we firstly introduce the system model of an asymmetric FD system with massive MIMO and the basic notation used in this paper.\footnote{Throughout this paper, we use upper- and lower-case boldfaces to denote matrix $\pmb{A}$ and vector $\pmb{a}$, respectively. We denote the inverse, transpose, and Hermitian of matrix $\pmb{A}$ as $\pmb{A}^{-1}$, $\pmb{A}^{T}$, and $\pmb{A}^{*}$, respectively. In addition, we denote the conjugate of a scalar $a$ as $a^*$. $\text{Var}[X]$ and $\mathbb{E}[X]$ respectively represent the variance and average of the random variable $X$, while $\text{Cov}(X,Y)$ represents the covariance between two random variables $X$ and $Y$. $\text{Sort}(X_k)$ is referred to as a sort function of sequence $X_k$ in the descending order.}   
	Then SINR, precoding/receive combining matrix design, and SIC methods are presented.

	\subsection{FD BS with Massive Arrays}
	We consider FD multi-user MIMO channels, as illustrated in Fig.~\ref{Sysmodel}, with a large number of antennas at the BS that serves $K_\text{DL}$ downlink users and $K_\text{UL}$ uplink users equipped with one antenna for each. 
	It is assumed that the BS is equipped with $M_\text{total}$ ($M_\text{total}>>K$~\cite{Lim_TWC}) RF chains that are divided into $M$ transmit and $N$ receive RF chains by considering their physical size, whereas each user has a single transmit and single receive RF chain.
	All users are assumed to operate in a half-duplex mode, so the hardware complexity for the SIC at the user is zero, which is beneficial for their battery uses. 
	This concept has been widely adopted in massive MIMO systems in cellular scenarios~\cite{Yin_asi13,Shojaeifard_TCOM17,Koh_TWC18}. 
	In addition, we assume an asymmetric data traffic load (i.e., the downlink has more data traffic than the uplink). In cellular systems, there hardly exists any user equipment (UE) requiring contents with high data rates for both links within the same coherence time. This stands in contrast to device-to-device scenarios. Thus, supporting the best-performing UE in a half-duplex mode can achieve greater scheduling gain than that in an FD mode. We consider that both the RF chain usage and data traffic load are asymmetric. Therefore, we assume $M\ge N+K_\text{DL}$ and $N \ge K_\text{UL}$. 
	This asymmetric antenna usage ensures a small and low-cost BS by reducing the receive RF chains that include low-noise amplifiers (LNA) within the fixed size of $M_\text{total}$. 
	Without loss of generality, let us consider that each propagation gain of the SI channels between BS antennas, denoted by $\beta^\text{SI}$, is the same as their expectation.
	In addition, the notations defined in this paper are shown in Table~\ref{Tab.Syms}.
	\\

	\begin{table}
	\caption{}
		\label{Tab.Syms}
		\begin{tabular}{p{3cm}|p{5cm}} \hline \hline
			\textbf{Symbol}        & \textbf{Description}         \\ \hline \hline
			\texttt{$M_\text{total}$(=$M+N$)}  & The number of RF chains (transmit RF chains $+$ receive RF chains)                   \\
			\texttt{$K_\text{DL}$/ $K_\text{UL}$}  & The number of downlink/ uplink users                    \\
			\texttt{$\rho_\text{d}(=\text{P}_\text{d}/\text{P}_\text{n})$}  & The total transmit signal-to-noise ratio (SNR) at the BS                   \\
			\texttt{$\rho_{\text{u}}(=\text{P}_\text{u}/\text{P}_\text{n})$}  & The transmit SNR at the user                   \\
			\texttt{$\rhoDL(=\rho_\text{d}\betaDL)$}       & The received downlink SNR at the users                   \\ 
			\texttt{$\rhoUU(=\rho_\text{u}\betaUU)$}       & The received SI SNR at the users                   \\ 
			\texttt{$\rhoUL(=\rho_\text{u}\betaUL)$}  & The received uplink SNR at the BS                   \\
			\texttt{$\rho^{\text{SI}}(=\rho_\text{d}\beta^\text{SI})$}  & The received SI SNR at the BS                   \\
			\texttt{$\text{P}_\text{d}$, $\text{P}_\text{u}$, $\text{P}_\text{n}$}  & The total transmit power, the transmit power of the user, the noise power of the additive white Gaussian noise (AWGN) of the channel                   \\
			\texttt{$\betaDL$/ $\betaUL$}  & The propagation gain between the BS and the $k_\text{d}$-th DL user / the $k_\text{u}$-th UL user                   \\
			\texttt{$\betaUU$}  & The propagation gain from the $k_\text{u}$-th user to the $k_\text{d}$-th user                   \\
			\texttt{$\pmb{H}$/ $\pmb{R}$/ $\pmb{E}$}  & The channel/ correlation/ estimated error matrix  	\\
			\texttt{$\tau$}  & The number of pilot symbols for the channel  	\\
			\texttt{$\epsilon$}  & The variance of the estimation error for the channel 	\\
			\texttt{$\alpha_\text{anc}$}  & The level of passive analog SIC		\\
			\texttt{$\alpha_\text{tx}$, $\pmb{n}_\text{tx}$}  & The transmitter noise variance, the transmitter noise vector
			 	\\
			\texttt{$\pmb{n}_\text{q}$, $\pmb{n}_\text{p}$}  & The additive quantization noise vector, the AWGN vector at SI
			\\
			\texttt{$\pmb{n}_{_\text{DL}}$/ $\pmb{n}_{_\text{UL}}$}  & The AWGN vector at downlink/ uplink
			 \\
			\texttt{$s$/ $x$}  & The transmit symbol at downlink/ uplink
			\\
			\texttt{$\pmb{G}$, $\pmb{W}$}  & The transmit precoding matrix, the receive combining matrix
			\\	
			\texttt{$\gamma$}  & The SINR
			\\	
			\texttt{$\Omega$}  & The SI power term
			\\			
			\hline
		\end{tabular}
	\end{table}

	
	
	

	\subsection{Channel Model}
	\subsubsection{Spatially Correlated Channels}
	
	The downlink, uplink, and SI channels are modeled by the $K_\text{DL}\times M$ matrix $\pmb{H}_\text{DL}=\pmb{R}_\text{RX(DL)}^{1/2}\pmb{H}_\text{DL,iid}\pmb{R}_\text{TX(DL)}^{1/2}$, $N\times K_\text{UL}$ matrix $\pmb{H}_\text{UL}=\pmb{R}_\text{RX(UL)}^{1/2}\pmb{H}_\text{UL,iid}\pmb{R}_\text{TX(UL)}^{1/2}$, and $N\times M$ matrix $\pmb{H}_\text{SI}=\pmb{R}_\text{RX(SI)}^{1/2}\pmb{H}_\text{SI,iid}\pmb{R}_\text{TX(SI)}^{1/2}$, which are based on the Kronecker model.
	The element of $\pmb{R}_\text{TX}$ and $\pmb{R}_\text{RX}$ is $r_\text{TX/RX}^{ij} = J_0(2\pi{d_\text{TX/RX}^{ij}}/\lambda)$, where $d_\text{TX/RX}^{ij}$ is the distance between the $i$-th array and the $j$-th array of the transmit/receive antenna at the BS, $J_0$ and $\lambda$ are a zero-order Bessel function of the first kind and a carrier wavelength, respectively.
	The specific parameters used in the simulation are presented in Section~\ref{Section:C3_Section5}.
	The elements of $\pmb{H}_\text{DL,iid}$, $\pmb{H}_\text{UL,iid}$, and $\pmb{H}_\text{SI,iid}$ are i.i.d. complex Gaussian random variables with zero mean and unit variance. The transmit correlation matrices in the downlink, uplink and SI channels are denoted as $\pmb{R}_\text{TX(DL)}$, $\pmb{R}_\text{TX(UL)}$, and $\pmb{R}_\text{TX(SI)}$, respectively, while the receive correlation matrices in these channels are denoted as $\pmb{R}_\text{RX(DL)}$, $\pmb{R}_\text{RX(UL)}$, and $\pmb{R}_\text{RX(SI)}$, respectively. It should be noted that, in contrast to the uplink and downlink channels, the SI channel is highly correlated in the spatial domain. Despite the high level of analog SIC, the power from direct paths (which are line-of-sight (LOS) paths, reflected paths around the antenna, and circuit leakage) is still greater than the power from non-line-of-sight (NLOS) paths. In addition, because the angular spread of the LOS path is very small, the spatial correlation of the SI channel is high. Thus, it is assumed that the SI channel is more highly correlated than the downlink and uplink channels. Furthermore, because the SI channel has a strong LOS component (a Rician channel model), a highly correlated channel and a real SI channel (which are equivalent channel models with a strong LOS component) are considered to supplement the SI channel modeling.

	\subsubsection{MMSE-Based Channel Estimation}
	
	Let $\pmb{\hat H}_{\text{DL}}$, $\pmb{\hat H}_{\text{UL}}$, and $\pmb{\hat H}_{\text{SI}}$ be the estimated channel matrices of $\pmb{H}_{\text{DL}}$, $\pmb{H}_{\text{UL}}$, and $\pmb{H}_{\text{SI}}$, respectively.
	Because estimation errors are independent of the original channels, the estimated channels can be modeled using estimated error matrices ($\pmb{E}$) that have i.i.d. elements with the distribution $\mathcal{CN}(0,\epsilon^2)$ as follows:
	$
	\pmb{\hat H}_\text{DL/UL/SI}=\pmb{H}_\text{DL/UL/SI}+\pmb{E}_\text{DL/UL/SI}.\label{est_error}
	$
	It is assumed that an MMSE-based channel estimation is used at the BS, which is also used in the 3rd Generation Partnership Project (3GPP)~\cite{3GPP_CHE}. 
	Other estimation schemes, such as least square-based estimation, can also be adapted to this scheme~\cite{Ngo_TCOM12}.
	In practice, the estimation error of the SI channel is affected by transmitter noise caused by hardware impairments\footnote{We consider the transmitter noise only in the SI channel, because its level is higher than that of the receiver noise~\cite{Day_TSP}. In the downlink and uplink channels, this noise can be negligible owing to the large path loss. In addition, we disregard the receiver distortion resulted from the hardware impairment at the receiver because it is much lower than the receiver noise.}, such as the nonlinearity of the power amplifier, a quantization error, and phase noise \cite{Day_TSP}.
	Thus, it is difficult to model the estimated SI channel while taking into consideration hardware impairments. The variance of the estimation error is simply modeled as the normalized-MSE (NMSE) value in~(\ref{est_error}) between the estimated channel and the original channel. This error depends on the performance of the estimation algorithm and the specifications of the hardware. 
	Because we consider a scenario of ADCs with a few bits (e.g., 4-5 bits) for realizing a cost-effective design~\cite{Bai_ISWCS13, Orhan_ITA15, Fan_COML15}, the variance of the estimation error of the SI channel is given by $\epsilon_{_\text{SI}}^2=\text{NMSE}$, which is derived from the AQNM as shown in~\ref{CE_error}.

	\begin{remark}
		It should be noted that the variance of the estimation error of the SI channel can also be modeled by the number of pilot symbols. Let the pilot symbols be transmitted orthogonally under TDD-based transmission. The numbers of pilot symbols for the SI channels are given by $\tau_{_\text{SI}}(\ge N)$ ($N$ transmit RF chains are connected to $N$ receive antennas through the switches).
		From \cite{Koh_TWC18}, the variance of the estimation error of the SI channel is given by
		\begin{align}
		\epsilon_{_\text{SI}}^2=\frac{\alpha_\text{tx}\rho_{\text{d}}N\frac{\beta^\text{SI}}{\alpha_\text{anc}}+1}  
		{\tau_{_\text{SI}}\rho_{\text{d}}\frac{\beta^\text{SI}}{\alpha_\text{anc}}+\alpha_\text{tx}\rho_{\text{d}}N\frac{\beta^\text{SI}}{\alpha_\text{anc}}+1}
		=\frac{\alpha_\text{tx}N\frac{\rho^\text{SI}}{\alpha_\text{anc}}+1}{\tau_{_\text{SI}}\frac{\rho^\text{SI}}{\alpha_\text{anc}}+\alpha_\text{tx}N\frac{\rho^\text{SI}}{\alpha_\text{anc}}+1},
		\end{align}
	\end{remark}
	where $\alpha_\text{anc}$ is the level of the passive analog SIC, and the transmitter noise is a complex Gaussian random variable with zero mean and smaller variance than the receiver noise (denoted by $\alpha_\text{tx}$ ($\alpha_\text{tx}\ll1$)).
	
	In the case of the downlink/uplink channel estimation, we assume there are no error terms arising from the hardware impairments. Thus, the variances of the estimation error for the downlink and uplink channels are given by~\cite{Ngo_TCOM12} 
	\begin{align}
	\epsilon_{_\text{DL},\kd}^2=\frac{1}{K_\text{DL}\rho_{\text{u}^\text{(DL)}}\beta^\text{DL}_{\kd}+1},~~
	\epsilon_{_\text{UL},\ku}^2=\frac{1}{K_\text{UL}\rho_{\text{u}}\beta^\text{UL}_{\ku}+1}\label{esterr_dlul}.
	\end{align}
	Here, the received pilot SNRs of the downlink and uplink users are $K_\text{DL}\rho_{\text{u}}^\text{(DL)}$ and $K_\text{UL}\rho_{\text{u}}$, respectively. In addition, the variances of the estimated channel for the $\kd$-th and $\ku$-th users are given by
	\begin{align}
	\text{Var}[\hat{h}^\text{DL}_{\kd m}]&=\frac{K_\text{DL}\rho_{\text{u}^\text{(DL)}}\beta^\text{DL}_{\kd}}{K_\text{DL}\rho_{\text{u}^\text{(DL)}}\beta^\text{DL}_{\kd}+1},\nonumber\\
	\text{Var}[\hat{h}^\text{UL}_{n \ku}]&=\frac{K_\text{UL}\rhoUL}{K_\text{UL}\rhoUL+1}\label{Var_UL_estChannel}.
	\end{align}
	If the received pilot SNRs are high (i.e., both SNRs approach infinity), one may consider perfect downlink and uplink channel estimations, which means $\text{Var}[\hat{h}^\text{DL}_{\kd m}]\rightarrow1$ and $\text{Var}[\hat{h}^\text{UL}_{n \ku}]\rightarrow1$, and these are the same as $\text{Var}[{h}^\text{DL}_{\kd m}]$ and $\text{Var}[{h}^\text{UL}_{n \ku}]$, respectively. It should be noted that any estimation algorithm dependent on the transmit pilot power can be modeled similarly.

	\subsubsection{Channel Estimation Error Model} \label{CE_error}
	
	We assume that each RF chain is connected to a $p$ bits ADC at the BS. To represent a quantization error in a channel estimation error, we use the AQNM~\cite{Orhan_ITA15, Fan_COML15} as follows:
	$
	\pmb{U}_\text{SIQ}=\alpha\pmb{U}_\text{SI}+\pmb{N}_\text{q},\label{AQNM}
	$
	where $\pmb{U}_\text{SIQ}$ is the $M\times N$ output matrix of the ADC corresponding to the $M\times N$ input matrix $\pmb{U}_\text{SI}$, which is the received SI signal at the BS. 
	Accordingly, $\alpha$ can be computed as $1-\frac{p^{-2}(\pi \sqrt{3})}{2}$. 
	Here, $\pmb{U}_\text{SI}$ can be modeled as~\cite{Ngo_TCOM12}
	\begin{align}
	\pmb{U}_\text{SI}=\sqrt{\frac{\rho _{\text{SI}}}{\alpha _{\text{anc}}}}\pmb{\text{H}}_{\text{SI}}^{T}(\pmb{P}+\pmb{N}_\text{tx})+\pmb{N}_\text{p},\label{ysi}
	\end{align}
	where $\pmb{N}_{\text{tx}}$, $\pmb{P}$, and $\pmb{N}_\text{p}$ are the transmitter noise, the $M\times N$ matrix that satisfies $\pmb{P}\pmb{P}^{H}=\pmb{I}$, and the $M\times N$ complex Gaussian random variables with i.i.d. elements following the distribution $\mathcal{CN}(0,\epsilon^2)$, respectively.
	$\pmb{N}_\text{q}$ is the $M\times N$ additive quantization noise matrix such that $\pmb{U}_\text{SI}$ and $\pmb{N}_\text{q}$ are uncorrelated. 
	It should be noted that the variance of $\pmb{N}_\text{q}$ is given by 		
	\begin{align}
	\text{Var}[\pmb{N}_\text{q}]&=\alpha(1-\alpha)\text{Var}[{\pmb{U}_\text{SI}}].\label{varNq}\nonumber\\
	&\overset{(2)}{=}\alpha(1-\alpha)(\left ( \frac{\rho _{\text{SI}}}{\alpha _{\text{anc}}}  \right )(1+\alpha_\text{tx})+1).
	\end{align}
	Thus, the MMSE estimate of $\pmb{\hat H}_\text{SI}$ given $\pmb{U}$ is
	$
	\widehat{\pmb{H}_\text{SI}}=(\sqrt{\frac{\alpha _{\text{anc}}}{\rho _{\text{d}}\beta_{\text{SI}} }}\pmb{\text{U}}_{\text{SIQ}}^{T}\pmb{P}\ast \widetilde{\pmb{D}})^{T},
	$
	where $\widetilde{\pmb{D}}=(\frac{\alpha _{\text{anc}}}{\rho _{\text{d}}}\beta_{\text{SI}}+I)^{-1}$.

	\subsection{SINR at Uplink and Downlink}
	Let $\pmb{g}_{\kd}$ and $s_{\kd}$ denote the $\kd$-th column vector of the transmit precoding matrix $\pmb{G}$ and the transmit symbol for the $\kd$-th user at the downlink, respectively. 
	Similarly, let $\pmb{w}_{\ku}$ denote the $\ku$-th column vector of the receive combining matrix $\pmb{W}$ and $x_{\ku}$ represent the transmit symbol for the $\ku$-th user at the uplink.
	In addition, let $\pmb{n}_{_\text{DL}}$ and $\pmb{n}_{_\text{UL}}$ be the additive white Gaussian receiver noise vector at the downlink and uplink, respectively. 
	The received signal at the $\kd$-th user is then expressed as
	\begin{align}
	y_{\kd}
	&=\underbrace{\sqrt{\rho_{_\text{DL}}}\pmb{\hat h}_{\text{DL}, \kd}^T\pmb{g}_{\kd}s_{\kd}}_{\text{desired signal}}+\underbrace{\sqrt{\rho_{_\text{DL}}}\sum_{\ell=1,\ell\neq \kd}^{K_\text{DL}}\pmb{\hat h}_{\text{DL}, \kd}^T\pmb{g}_{\ell}s_\ell}_{\text{user-interference}}\nonumber\\
	&~~~+\underbrace{\sqrt{\rho_{_\text{DL}}}\sum_{\ell=1}^{K_\text{DL}}\pmb{e}_{\text{DL}, \kd}^T\pmb{g}_{\ell}s_\ell}_{\text{estimation error}}
	+\underbrace{\sum_{\ku=1}^{K_\text{UL}}\sqrt{\rhoUU} h_{\kd\ku}^{\text{UU}}x_{\ku}}_{\text{user-to-user-interference}}+n_{_\text{DL}, \kd},\label{eq_rcv_dl}
	\end{align}
	where $\pmb{\hat h}_{\text{DL}, \kd}$ represents the $\kd$-th column vector of $\pmb{\hat H}_\text{DL}^T$ and $h_{\kd\ku}$ is the interference channel from the $\ku$-th user to the $\kd$-th user.
	The received signal for the $\ku$-th user at the BS is expressed as
	\begin{align}
	u_{\ku}
	&=\underbrace{\sqrt{\rhoUL}\pmb{w}_{\ku}^T\pmb{\hat h}_{\text{UL}, \ku}x_{\ku}}_{\text{desired signal}}
	+\underbrace{\sum_{\ell=1,\ell\neq {\ku}}^{K_\text{UL}}\sqrt{\rhoULell}\pmb{w}_{\ku}^T\pmb{\hat h}^{\text{UL}}_{ \ell}x_\ell}_{\text{user-interference}}\nonumber\\
	&~~~+\underbrace{\sum_{\ell=1}^{K_\text{UL}}\sqrt{\rhoULell}\pmb{w}_{\ku}^T\pmb{e}^{\text{UL}}_{ \ell}x_\ell}_{\text{estimation error}}
	+\underbrace{\sqrt{\rho^\text{SI}/\alpha_\text{anc}}\pmb{w}_{\ku}^T\pmb{H}_\text{SI}\pmb{G}\pmb{\bar{s}}}_{\text{self-interference}}+\pmb{w}_{\ku}^T\pmb{n}_{_\text{UL}},\label{eq_rcv_ul}
	\end{align}
	where $\pmb{\hat h}_{\text{UL}, \ku}$ is the $\ku$-th column vector of $\pmb{\hat H}_\text{UL}$. In addition, $\pmb{\bar{s}}$ represents $\pmb{s}+\pmb{n}_{_\text{tx}}$. To satisfy a power constraint, the normalized transmit beamforming vectors (the columns of the precoding matrix) with vector normalization are given as $\pmb{g}_{\kd}=\pmb{f}_{\kd}/(\sqrt{K_\text{DL}}||\pmb{f}_{\kd}||)$, where $1\le\kd\le K_\text{DL}$~\cite{Lim_TWC}.
	
	From (\ref{eq_rcv_dl}) and (\ref{eq_rcv_ul}), we can derive the SINRs of the $\kd$-th user at the downlink and the $\ku$-th user at the uplink, respectively, as follows:
	\begin{align}
	{\gamma}_{\kd}^\text{DL}&=\frac{\rhoDL\left|\pmb{\hat h}_{\text{DL}, \kd}^T\frac{\pmb{f}_{\kd}}{\sqrt{K_\text{DL}}||\pmb{f}_{\kd}||}\right|^2}{\mathcal{I}_\text{error}^\text{DL}+\mathcal{I}_\text{user}^\text{DL}+\sum_{\ku=1}^{K_\text{UL}}\rhoUU |h_{\kd\ku}^{\text{UU}}|^2+1},\nonumber\\
	{\gamma}_{\ku}^\text{UL}&=\frac{\rhoUL|\pmb{w}_{\ku}^T\pmb{\hat h}_{\text{UL}, \ku}|^2}{\mathcal{I}_\text{error}^\text{UL}+\mathcal{I}_\text{user}^\text{UL}+\frac{\rho^\text{SI}}{\alpha_\text{anc}}\Omega(1+\alpha_\text{tx})+||\pmb{w}_{\ku}||^2},\label{SINR_UL}
	\end{align}
	where $\mathcal{I}_\text{error}^\text{DL}=\rhoDL\sum_{\ell=1}^{K_\text{DL}}\left|\pmb{e}_{\text{DL}, \kd}^T\frac{\pmb{f}_{\ell}}{\sqrt{K_\text{DL}}||\pmb{f}_{\ell}||}\right|^2$, $\mathcal{I}_\text{user}^\text{DL}=\rhoDL\sum_{\ell=1,\ell\neq \kd}^{K_\text{DL}}\left|\pmb{\hat h}_{\text{DL}, \kd}^T\frac{\pmb{f}_{\ell}}{\sqrt{K_\text{DL}}||\pmb{f}_{\ell}||}\right|^2$, $\mathcal{I}_\text{error}^\text{UL}=\sum_{\ell=1}^{K_\text{UL}}\rhoULell|\pmb{w}_{\ku}^T\pmb{e}_{\text{UL}, \ell}|^2$, $\mathcal{I}_\text{user}^\text{UL}=\sum_{\ell=1,\ell\neq \ku}^{K_\text{UL}}\rhoULell|\pmb{w}_{\ku}^T\pmb{\hat h}_{\text{UL}, \ell}|^2$, and $\Omega=||\pmb{w}_{\ku}^T\pmb{H}_\text{SI}\pmb{G}||^2$.

	\subsection{Precoding/Receive Combining Matrix Design and SIC Methods} 
	Because the FD BS always perfectly knows its precoding/receive matrix and transmitted downlink symbols, it can mitigate SI by directly subtracting $\sqrt{\rho^\text{SI}/\alpha_\text{anc}}\pmb{w}_{\ku}^T\hat{\pmb{H}}_\text{SI}\pmb{G}\pmb{s}$, of which the SIC is referred to as the SI subtraction. The received signal after the SI subtraction is
	\begin{align}
	u_{\text{stt}, \ku}=u_{\ku}-\sqrt{\rho^\text{SI}/\alpha_\text{anc}}\pmb{w}_{\ku}^T\hat{\pmb{H}}_\text{SI}\pmb{G}\pmb{\bar{s}}.
	\end{align}
	Here, the subscript $\text{stt}$ indicates the SI subtraction. The ZF precoder for the downlink is assumed at the BS in the cases of the FD BS with the SI subtraction and without SIC:
	\begin{align}
	\pmb{G}_\text{ZF}=\pmb{\hat H}_\text{DL}^*(\pmb{\hat H}_\text{DL}\pmb{\hat H}_\text{DL}^*)^{-1}.
	\end{align}
	In addition, because the precoding/combining matrices are computed only once per coherence block, the complexity difference between MF and ZF/MMSE is relatively small (the bulk of the complexity comes from FFTs and matrix-vector multiplications performed on a per-symbol basis~\cite{mmimo_com}). In addition, because the FD massive MIMO considered has many antennas ($>>the\ number\ of\ users$) and a high SNR environment, both ZF and MMSE have similar performance. Furthermore, ZF can provide a clear performance comparison with SI subtraction to satisfy our main contribution. For these reasons, the ZF is considered for downlink transmission. Future work will consider various precoding/combining matrices, including MRT/MRC, generalized MMSE and so on. From the antenna configuration of $M\ge N+K$, additional antennas can be utilized for spatial suppression \cite{FDSI_TSP11, Yin_asi13,Shojaeifard_TCOM17}. Spatial suppression-based algorithms for massive MIMO systems were proposed in \cite{Yin_asi13,Shojaeifard_TCOM17}.
	We adopt the extended ZF in~\cite{Yin_asi13} for spatial suppression, which requires minimum computational complexity for nulling.
	\begin{align}
	\pmb{G}_\text{sps}=\pmb{\hat H}_\text{ext}^*(\pmb{\hat H}_\text{ext}\pmb{\hat H}_\text{ext}^*)^{-1},~~~~\pmb{\hat H}_\text{ext}=\begin{bmatrix}\pmb{\hat H}_\text{DL}\\ \pmb{\hat H}_\text{SI}\end{bmatrix}.
	\end{align}
	Here, the subscript $\text{sps}$ indicates the spatial suppression. It should be noted that the vector set, from the beginning of the column vector of $\pmb{G}_\text{sps}$ to the $K_\text{DL}$-th column vector, is the same as the null-space projection precoder in \cite{FDSI_TSP11}, where the rank of $\pmb{H}_\text{SI}$ is $N$, and the nullity of that is $K_\text{DL}$.
	
	To eliminate user-interference signals at the uplink, we use the following ZF filter in both the transceiver with the SI subtraction/spatial suppression and without the SIC:
	\begin{align}
	\pmb{W}=({\pmb{\hat H}}_\text{UL}^*{\pmb{\hat H}}_\text{UL})^{-1}{\pmb{\hat H}^*_\text{UL}}.
	\end{align}

	\section{Performance Comparison in Different Channel Conditions}
		\label{sec:channel}
	
	In this section, we compare the performance of two SIC methods with channel conditions which are correlated/uncorrelated and perfect/imperfect. Specifically, we provide mathematical motivations and preliminaries at first. Then, based on these preliminaries, we compare the two methods by SINR or the expected SI power term in uncorrelated channels. Finally, in subsection~\ref{correl}, we compare them by the expected SI power term in correlated channels.
	
	\subsection{Mathematical Motivations and Preliminaries}
	
	In this subsection, we provide mathematical motivations and preliminaries to investigate the ergodic sum rate and SI power for FD massive MIMO systems. We also present an asymptotic analysis of FD massive MIMO systems under the perfect channel estimation conditions.
	
	\subsubsection{SI Power Term in Massive MIMO Systems}
	In the FD system, the SI level degrades the uplink performance as a bottleneck, which is represented as the SI power term $\Omega$ in (\ref{SINR_UL}). Let $\pmb{A}$ be an $N\times M$ random matrix, the elements of which are complex random variables with zero mean and $\sigma_A^2$. In addition, $w_{\ku n}$, $a_{nm}$, and $g_{m\ell}$ are the $n$-th elements of $\pmb{w}_{\ku}$, the $(n, m)$-th element of $\pmb{A}$, and the $(m, \ell)$-th element of the $M\times L$ matrix $\pmb{G}$ (the precoder matrix), respectively. Then, the quadratic form of $||\pmb{w}_{\ku}^T\pmb{A}\pmb{G}||^2$ can be expressed as 
	\begin{align}
	||\pmb{w}_{\ku}^T\pmb{A}\pmb{G}||^2=\sum_{m=1}^L\sum_{i=1}^N\sum_{j=1}^N\sum_{k=1}^M\sum_{\ell=1}^M w_{\ku i}w_{\ku j}^*a_{ik}a_{j\ell}^*g_{km}g_{\ell m}^*.\label{eqquad}
	\end{align}
	It should be noted that the SI power term can be derived by replacing $\pmb{A}$ in (\ref{eqquad}) with $\pmb{H}_\text{SI}$ or $\pmb{E}_\text{SI}$ according to the SIC method.
	
	\begin{lemma}\label{decomposition}
		\emph{(decomposition of the expected SI power term)} The expectation of (\ref{eqquad}) can be decomposed into
		\begin{align}
		&\mathbb{E}\left\{||\pmb{w}_{\ku}^T\pmb{A}\pmb{G}||^2\right\}\nonumber
		\\&=\sum_{m=1}^L\sum_{i=1}^N\sum_{j=1}^N\sum_{k=1}^M\sum_{\ell=1}^M\mathbb{E}\left\{ w_{\ku i}w_{\ku j}^*a_{ik}a_{j\ell}^*g_{km}g_{\ell m}^*\right\}\nonumber\\
		&\overset{(a)}{=}\sum_{m=1}^L\sum_{i=1}^N\sum_{j=1}^N\sum_{k=1}^M\sum_{\ell=1}^M\mathbb{E}\left\{ w_{\ku i}w_{\ku j}^*\right\}\mathbb{E}\left\{a_{ik}a_{j\ell}^*g_{km}g_{\ell m}^*\right\}\nonumber
		\end{align}
		\begin{align}
		&=\underbrace{\sum_{m=1}^L\sum_{i=1}^N\sum_{k=1}^M\sum_{\ell=1}^M\mathbb{E}\left\{ |w_{\ku i}|^2\right\}\mathbb{E}\left\{a_{ik}a_{i\ell}^*g_{km}g_{\ell m}^*\right\}}_{\text{expectation term related to i.i.d. $\pmb{H}_{UL}$}}+\widetilde{\Omega}_{\pmb{A}(i\neq j)}\nonumber\\
		&=\sum_{i=1}^N\mathbb{E}\left\{ |w_{\ku i}|^2\right\} \nonumber
		\\&~~\times\sum_{m=1}^L\sum_{k=1}^M\left[\underbrace{\mathbb{E}\left\{|a_{ik}|^2\right\}\mathbb{E}\left\{|g_{km}|^2\right\}}_{\text{expectation term where $\pmb{A}$ and $\pmb{G}$ are independent}}+\zeta\right]\nonumber
		\\&~~+\widetilde{\Omega}_{\pmb{A}(i\neq j)},\label{eq:decompSI}
		\end{align}
		where 
		\begin{align}
		&\widetilde{\Omega}_{\pmb{A}(i\neq j)}\nonumber
		\\&=\underbrace{\sum_{m=1}^L\sum_{i=1}^N\sum_{j=1,j\neq i}^N\sum_{k=1}^M\sum_{\ell=1}^M\mathbb{E}\left\{ w_{\ku i}w_{\ku j}^*\right\}\mathbb{E}\left\{a_{ik}a_{j\ell}^*g_{km}g_{\ell m}^*\right\}}_{\text{residual expectation term due to correlated $\pmb{H}_{UL}$}},\nonumber
		\end{align}
		\begin{align}
		\zeta=\underbrace{\text{Cov}(|a_{ik}|^2,|g_{km}|^2)+\sum_{\ell=1,\ell\neq k}^{M}\mathbb{E}\left\{a_{ik}a_{i\ell}^*g_{km}g_{\ell m}^*\right\}}_{\text{residual expectation term due to correlation between $\pmb{A}$ and $\pmb{G}$}}.\nonumber
		\end{align}
		Equation (\emph{a}) is derived based on the fact that $w_{\ku n}$ is independent of both $a_{nm}$ and $g_{m\ell}$. 
	\end{lemma}
	
	This lemma implies that the expected SI power varies according to the correlation between $\pmb{A}$ and $\pmb{G}$ as well as the correlation among their own elements.\footnote{The right side of $\zeta$ includes the terms that occurred due to the correlated $\pmb{G}$. We discuss this in \emph{Lemma}~\ref{lemma:correlatedchannel}.} Using this lemma, we compare the two SIC methods in correlated channel models. It should be noted that, the conventional analysis does not focus on these correlation terms by assuming independent conditions~\cite{Yin_asi13,Min_TVT16}.

	\begin{lemma}\label{corollary:ESIpower}
		\emph{(expected SI power term with independent conditions)} The following equation holds when $\pmb{w}_{\ku}$, $\pmb{A}$, and $\pmb{G}$ are independent of each other and, consist of i.i.d. random variables~\cite{Yin_asi13,Min_TVT16}:
		\begin{align}
		\mathbb{E}\left\{||\pmb{w}_{\ku}^T\pmb{A}\pmb{G}||^2\right\}=\sigma_A^2\mathbb{E}\left\{||\pmb{w}_{\ku}||^2\right\}.\nonumber
		\end{align}
		\begin{IEEEproof}
			See Appendix A.
		\end{IEEEproof}
	\end{lemma}
	
	\subsubsection{Approximation of Ergodic Sum Rate of Massive MIMO Systems}
	
	\begin{remark}
		Let $S$, $I$, and $N$ be the norms of the random vectors of the desired signal power, interference power, and noise power terms, respectively. From \cite[Lemma~4]{Lim_TWC}, the ergodic achievable sum rate of massive MIMO systems in the low/high SNR regime can be approximated as follows:
		\begin{align}
		\mathbb{E}\left\{\log_2\left\{1+\frac{S}{I+N}\right\}\right\}\approx\mathcal{R}=\log_2\left\{1+\frac{\mathbb{E}\{S\}}{\mathbb{E}\{I+N\}}\right\},\label{approxrate}
		\end{align}
		where $\mathbb{E}\{S\}\mathbb{E}\left\{\frac{1}{I+N}\right\}\approx\frac{\mathbb{E}\{S\}}{\mathbb{E}\{I+N\}}$ or $\mathbb{E}\left\{\frac{S}{I+N}\right\}\approx\frac{\mathbb{E}\{S\}}{\mathbb{E}\{I+N\}}$. From (\ref{approxrate}), we let $\bar{\gamma}$ $(=\frac{\mathbb{E}\{S\}}{\mathbb{E}\{I+N\}})$ be an approximation of $\gamma$.\label{remark_rate}
	\end{remark}

	\subsubsection{Performance Analysis with Perfect Channel Estimation} \label{perfect}

First, we begin with a brief overview of the ergodic achievable sum rate of the two SIC schemes with perfect channel state information ($\epsilon_{_\text{DL/UL/SI}}$= 0) in spatially uncorrelated channels.
In summary, from \emph{Remark}~\ref{remark_rate}, the approximations of the ergodic rate of the $\kd$-th user with the SI subtraction and the spatial suppression at downlink are expressed, respectively, as follows:  
\begin{align}
	\mathcal{R}_{\kd, \text{stt}}^\text{DL}
	&=\log_2\left\{1+\frac{\frac{\rhoDL}{K_\text{DL}}\mathbb{E}\left\{\frac{1}{||\pmb{f}_{\kd, \text{ZF}}||^2}\right\}}{\sum_{\ku=1}^{K_\text{UL}}\rhoUU \mathbb{E}\{|h_{\kd\ku}^{\text{UU}}|^2\}+1}\right\}\nonumber
	\\&=\log_2\left\{1+\frac{\rhoDL(M-K_\text{DL}+1)}{K_\text{DL}(\sum_{\ku=1}^{K_\text{UL}}\rhoUU+1)}\right\},\label{RDL_perfect1}
\end{align}
\begin{align}
	\mathcal{R}_{\kd, \text{sps}}^\text{DL}
	&=\log_2\left\{1+\frac{\frac{\rhoDL}{K_\text{DL}}\mathbb{E}\left\{\frac{1}{||\pmb{f}_{\kd, \text{sps}}||^2}\right\}}{\sum_{\ku=1}^{K_\text{UL}}\rhoUU \mathbb{E}\{|h_{\kd\ku}^{\text{UU}}|^2\}+1}\right\}\nonumber
	\\&=\log_2\left\{1+\frac{\rhoDL(M-N-K_\text{DL}+1)}{K_\text{DL}(\sum_{\ku=1}^{K_\text{UL}}\rhoUU+1)}\right\}.\label{RDL_perfect2}
\end{align}
In addition, the approximations of the ergodic rate of the $\ku$-th user with both SIC methods at the uplink are the same and are given by
\begin{align}
	\mathcal{R}_{\ku, \text{stt/sps}}^\text{UL}
	&=\log_2\left\{1+\mathbb{E}\left\{\frac{\rhoUL}{||\pmb{w}_{\ku}||^2}\right\}\right\}\nonumber
	\\&=\log_2\left\{1+\rhoUL(N-K_\text{UL}+1)\right\}\label{RUL_perfect2},
\end{align}
where $\mathcal{I}_\text{error}$, $\mathcal{I}_\text{user}=0$, $\mathbb{E}\left\{\frac{1}{||\pmb{f}_{\kd,\text{ZF}}||^2}\right\}=M-K_\text{DL}+1$, $\mathbb{E}\left\{\frac{1}{||\pmb{f}_{\kd,\text{sps}}||^2}\right\}=M-(N+K_\text{DL})+1$, and $\mathbb{E}\left\{\frac{1}{||\pmb{w}_{\ku}||^2}\right\}=N-K_\text{UL}+1$ (results are derived from the diversity order of ZF~\cite{Lim_TWC,diversity_order}).
From (\ref{RDL_perfect1})-(\ref{RUL_perfect2}), we obtain 
\begin{align}
	\mathcal{R}_\text{stt}&=\sum_{\kd=1}^{K_\text{DL}}\log_2\left\{1+\frac{\rhoDL(M-K_\text{DL}+1)}{K_\text{DL}(\sum_{\ku=1}^{K_{UL}}\rhoUU+1)}\right\}\nonumber
	\\&~~+\sum_{\ku=1}^{K_\text{UL}}\log_2\left\{1+\rhoUL(N-K_\text{UL}+1)\right\},
	\label{rate_stt}
\end{align}
\begin{align}
	\mathcal{R}_\text{sps}&=\sum_{\kd=1}^{K_\text{DL}}\log_2\left\{1+\frac{\rhoDL(M-N-K_\text{DL}+1)}{K_\text{DL}(\sum_{\ku=1}^{K_{UL}}\rhoUU+1)}\right\}\nonumber
	\\&~~+\sum_{\ku=1}^{K_\text{UL}}\log_2\left\{1+\rhoUL(N-K_\text{UL}+1)\right\}.
	\label{rate_sps}
\end{align}
It should be noted that the approximations of the ergodic sum rate of the half-duplex with ZF precoding and ZF receive combining is the same as $\frac{1}{2}\mathcal{R}_\text{stt}$ with $\rhoUU=0$ $(\forall\kd,\forall\ku)$~\cite{Lim_TWC}.

From (\ref{rate_stt}) and (\ref{rate_sps}), the sum rate of the SI subtraction is, in the Genie-aided system, always better than that of the spatial suppression owing to the same performance at the uplink and the performance degradation at the downlink. This is because the SI subtraction and spatial suppression transceivers can perfectly cancel out the SI at the uplink, and the spatial suppression requires an additional dimension for the nulling of the ZF precoder at the downlink. 
	
	\subsection{Asymptotic Rate in Uncorrelated Channels} 
	In this subsection, we present the asymptotic analysis of FD massive MIMO systems according to the two SIC methods: SI subtraction and spatial suppression. First, we assume uncorrelated channels for simplifying mathematical expressions to obtain insights into how the imperfect channel estimation affects the performance of the two SIC schemes.

	
	In the FD system, the SI level degrades the uplink performance as a bottleneck, which is represented as the SI power term $\Omega$ in (\ref{SINR_UL}).
	Let ${\bar \Omega}=\mathbb{E}\{\Omega\}$ be the expected SI power. If there is no SIC at the BS ($\alpha_\text{anc}=0$~dB), from \emph{Lemma}~\ref{corollary:ESIpower}, the expected SI power is given by
	\begin{align}
	{\bar \Omega}_\text{woSIC}&=\mathbb{E}\left\{||\pmb{w}_{ku}^T\pmb{H}_\text{SI}\pmb{G}||^2\right\}
	=\mathbb{E}\left\{||\pmb{w}_{\ku}||^2\right\}\mathbb{E}\left\{|h^\text{SI}_{n,m}|^2\right\}\nonumber
	\\&=\mathbb{E}\left\{||\pmb{w}_{\ku}||^2\right\}.
	\label{eq:eSIpower}
	\end{align}
	Similarly, because $\pmb{E}_\text{SI}$ and $\pmb{G}_\text{ZF}$ are independent, from (\ref{eq:decompSI}), the quadratic form of the expected SI power of the SI subtraction is given~by
	\begin{align}
	{\bar \Omega}_{\text{stt}}
	&=\mathbb{E}\left\{||\pmb{w}_{\ku}^T(\pmb{H}_\text{SI}-\pmb{\hat H}_\text{SI})\pmb{G}_\text{ZF}||^2\right\}
	=\mathbb{E}\left\{||\pmb{w}_{\ku}^T\pmb{E}_\text{SI}\pmb{G}_\text{ZF}||^2\right\}\nonumber\\&=\epsilon_{_\text{SI}}^2\mathbb{E}\left\{||\pmb{w}_{\ku}||^2\right\}.\label{stt_SI}
	\end{align}
	In contrast to the SI subtraction, $\pmb{E}_\text{SI}$ and $\pmb{G}_\text{sps}$ are dependent when spatial suppression is used, because $\pmb{G}_\text{sps}$ is a function of $\pmb{\hat H}_\text{SI}=\pmb{H}_\text{SI}+\pmb{E}_\text{SI}$.
	Thus, from \emph{Lemma}~\ref{decomposition}, we can derive the quadratic form of the expected SI power of the spatial suppression as follows:
	\begin{align}
	{\bar \Omega}_{\text{sps}}
	&=\mathbb{E}\left\{||\pmb{w}_{\ku}^T\pmb{H}_\text{SI}\pmb{\bar G}_\text{sps}||^2\right\}
	=\mathbb{E}\left\{||\pmb{w}_{\ku}^T(\pmb{\hat H}_\text{SI}-\pmb{E}_\text{SI})\pmb{\bar G}_\text{sps}||^2\right\}\nonumber
	\\&=\mathbb{E}\left\{||\pmb{w}_{\ku}^T\pmb{E}_\text{SI}{\pmb{\bar G}}_\text{sps}||^2\right\}\nonumber\\
	&=\sum_{i=1}^N\mathbb{E}\left\{|w_{\ku i}|^2\right\}\sum_{m=1}^{K_\text{DL}}\sum_{k=1}^M\mathbb{E}\left\{|E_{ik}|^2\right\}\mathbb{E}\left\{|g_{km, \text{sps}}|^2\right\}\nonumber
	\\&~~+\sum_{i=1}^N\mathbb{E}\left\{|w_{\ku i}|^2\right\}\sum_{m=1}^{K_\text{DL}}\sum_{k=1}^M\zeta_\text{sps}\nonumber\\
	&=\epsilon_{_\text{SI}}^2\mathbb{E}\left\{||\pmb{w}_{\ku}||^2\right\}
	+\sum_{i=1}^N\mathbb{E}\left\{|w_{\ku i}|^2\right\}\sum_{m=1}^{K_\text{DL}}\sum_{k=1}^M\zeta_\text{sps},\label{sps_SI}
	\end{align}
	where $\widetilde{\Omega}_{\pmb{A}(i\neq j)}=0$, $\zeta_{\text{sps}}
	=\text{Cov}(|E_{ik}|^2,|g_{km,\text{sps}}|^2)+\sum_{\ell=1,\ell\neq k}^{M}\mathbb{E}\left\{E_{ik}E_{i\ell}^*g_{km,\text{sps}}g_{\ell m,\text{sps}}^*\right\}\!$, and $E_{nm}$ is the $(n, m)$-th element of $\pmb{E}_\text{SI}$. 
	In addition, ${\pmb{\bar G}}_\text{sps}$ is the subset from the first to the $K_\text{DL}$-th columns of $\pmb{G}_\text{sps}$ and $\pmb{\hat H}_\text{SI}{\pmb{\bar G}}_\text{sps}=\pmb{0}$.
	

	\begin{lemma}\label{proposition:zetaineq}
		The residual expectation term in $\bar{\Omega}_\text{sps}$ owing to the correlation between $\pmb{E}_\text{SI}$ and $\pmb{\bar G}_\text{sps}$ is negative:

		\begin{align}
		\zeta_{\text{sps}}&=\text{Cov}(|E_{ik}|^2,|g_{km,\text{sps}}|^2)\nonumber
		\\&~~+\sum_{\ell=1,\ell\neq k}^{M}\mathbb{E}\left\{E_{ik}E_{i\ell}^*g_{km,\text{sps}}g_{\ell m,\text{sps}}^*\right\}\nonumber
		\\&~~\le 0.\nonumber
		\end{align}
		
		\begin{IEEEproof}
			See Appendix B.
		\end{IEEEproof}
	\end{lemma}
	
	\begin{figure}[t]
		\centerline{\resizebox{1\columnwidth}{!}{\includegraphics{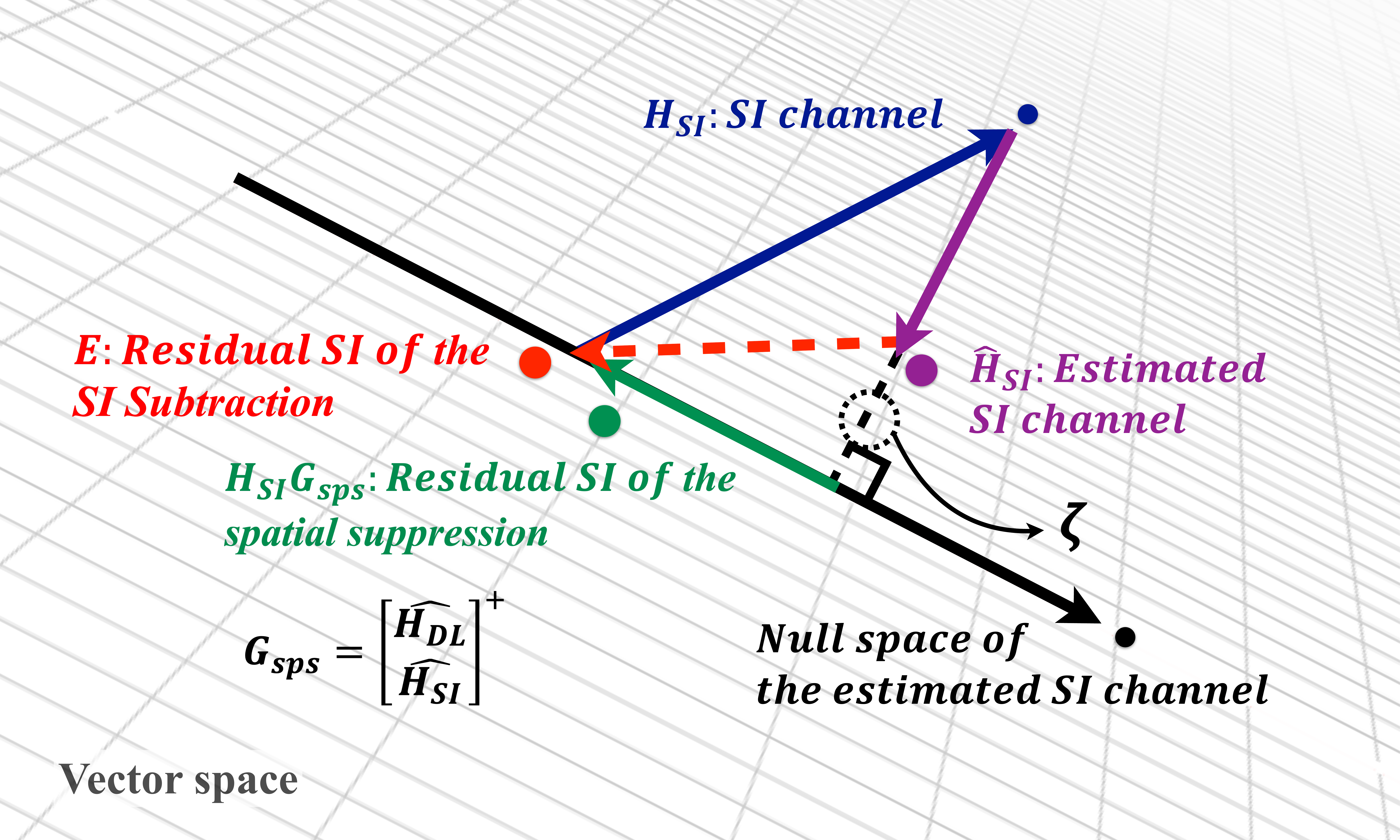}}}
		\caption{Concepts of digital SIC schemes in the vector space.}
		\label{Figure:nullspcace}
		\vspace{-14pt}
	\end{figure}
	
	By applying \emph{Lemma}~\ref{proposition:zetaineq} to the quadratic forms of $\bar{\Omega}_\text{stt}$ and $\bar{\Omega}_\text{sps}$ ((\ref{stt_SI}), (\ref{sps_SI})), we can confirm the inequality
	\begin{align}
	\bar{\Omega}_\text{stt}\ge\bar{\Omega}_\text{sps}.\label{inequality_imp}
	\end{align}
	
	To illustrate the implicit meaning of \emph{Lemma}~\ref{proposition:zetaineq}, we present Fig.~\ref{Figure:nullspcace}, which illustrates an example of the effect of spatial suppression on the residual SI by representing a manifold as a Euclidean space. This shows that, with imperfect channel estimation, the difference vector between the SI channel vector and their estimated vector (i.e., the estimation error vector) is additionally projected onto the null vector of the estimated SI channel vector. Thus, the length of the residual SI vector of the spatial suppression can be shorter than that of SI subtraction. Therefore, the expected SI power of the spatial suppression can be mitigated to a greater extent than that of SI subtraction.
	
	It is trivial from (\ref{rate_stt}) and (\ref{rate_sps}) that the SI subtraction outperforms the spatial suppression in the perfect channel estimation condition, i.e., $\mathcal{R}_\text{stt}\ge\mathcal{R}_\text{sps}$. However, as the ergodic uplink performance of the spatial suppression is the same as that of the SI subtraction in the perfect channel estimation condition, it is better than that of the SI subtraction as determined from~(\ref{inequality_imp}), i.e., $\bar{\Omega}_\text{stt}\ge\bar{\Omega}_\text{sps}$.

	Next, we present some approximations of the uplink SINR for each SIC method to support the obtained results.
	Using the approximation methods of \cite{Ngo_TCOM12} and \emph{Remark}~\ref{remark_rate}, the approximation of the uplink SINR without the SIC can be derived from (\ref{esterr_dlul}), (\ref{SINR_UL}), and ${\bar \Omega_{\text{woSIC}}}=\mathbb{E}\left\{||\pmb{w}_{\ku}||^2\right\}$ as follows:
		\begin{align}
		&\bar{\gamma}_{\text{woSIC}, \ku}^\text{UL}\nonumber
		\\&=\frac{\rhoUL}{\mathbb{E}\{||\pmb{w}_{\ku}||^2\}\sum_{\ell=1}^{K_\text{UL}}\frac{\rhoULell}{K_\text{UL}\rhoULell+1}+AA+\mathbb{E}\{||\pmb{w}_{\ku}||^2\}}\nonumber\\
		&=\frac{(K_\text{UL}\rhoUL+1)(N-K_\text{UL}+1)}{K_\text{UL}(\sum_{\ell=1}^{K_\text{UL}}\frac{\rhoULell}{K_\text{UL}\rhoULell+1}+1)+\frac{\rho^{\text{SI}}}{\alpha_\text{anc}}K_\text{UL}(1+\alpha_\text{tx})}\label{SINR_noSIC},
		\end{align}
		where $AA=\frac{\rho^{\text{SI}}}{\alpha_\text{anc}}\mathbb{E}\{||\pmb{w}_{\ku}||^2\}(1+\alpha_\text{tx})$, $|\pmb{w}_{\ku}^T\pmb{\hat h}_{\text{UL}, \ku}|^2=1$, $|\pmb{w}_{\ku}^T\pmb{\hat h}_{\text{UL}, \ell}|^2=0$, and $\mathbb{E}\left\{\frac{1}{||\pmb{w}_{\ku}||^2}\right\} = \frac{K_\text{UL}\rhoUL+1}{K_\text{UL}\rhoUL}(N-K_\text{UL}+1)$, which is calculated from~(\ref{Var_UL_estChannel}).

	Similarly, from (\ref{SINR_UL}) and (\ref{stt_SI}), we can derive the approximation of the uplink SINR with the SI subtraction as follows
	\begin{align}
	\bar{\gamma}_{\text{stt}, \ku}^\text{UL}
	=\frac{(K_\text{UL}\rhoUL+1)(N-K_\text{UL}+1)}{K_\text{UL}(\sum_{\ell=1}^{K_\text{UL}}\frac{\rhoULell}{K_\text{UL}\rhoULell+1}+1)+\frac{\rho^{\text{SI}}}{\alpha_\text{anc}}\epsilon_{_\text{SI}}^2K_\text{UL}(1+\alpha_\text{tx})}.\label{SINR_stt}
	&
	\end{align}
	
	Finally, we present the uplink SINR for spatial suppression.
	To obtain the independent condition, we should consider, given that $\pmb{H}_\text{SI}$ and $\pmb{G}_\text{sps}$ are not independent of each other, the best and worst estimation errors to obtain the independent condition. In the case of the best channel estimation error ($\epsilon_{_\text{SI}}\rightarrow0$), the estimated SI channel tends to the original SI channel ($\pmb{\hat H}_{\text{SI}}\rightarrow\pmb{H}_{\text{SI}}$), and thus $\pmb{E}_\text{SI}$ and $\pmb{G}_\text{sps}$ tend to be independent. Accordingly, from \emph{Lemma}~\ref{corollary:ESIpower}, the expected SI power of the spatial suppression (the best case) is given by
	\begin{align}
	{\bar \Omega}_{\text{sps}(\epsilon_{_\text{SI}}\rightarrow0)}
	&=\mathbb{E}\left\{||\pmb{w}_{\ku}^T\pmb{H}_\text{SI}\bar{\pmb{G}}_\text{sps}||^2\right\}\nonumber\\
	&=\mathbb{E}\left\{||\pmb{w}_{\ku}^T(\pmb{\hat H}_\text{SI}-\pmb{E}_\text{SI})\bar{\pmb{G}}_\text{sps}||^2\right\}\nonumber\\
	&=\mathbb{E}\left\{||\pmb{w}_{\ku}^T\pmb{E}_\text{SI}\bar{\pmb{G}}_\text{sps}||^2\right\}\nonumber\\
	&=\epsilon_{_\text{SI}}^2\mathbb{E}\left\{||\pmb{w}_{\ku}||^2\right\}\label{sps_SI_b}.
	\end{align}
	Similarly, in the case of the worst channel estimation error ($\epsilon_{_\text{SI}}\rightarrow\infty$), the estimated SI channel tends to the channel estimation error matrix ($\pmb{\hat H}_{\text{SI}}\rightarrow\pmb{E}_\text{SI}$) and $\pmb{\hat H}_\text{ext}$ tends to $\begin{bmatrix}\pmb{\hat H}_\text{DL}\\ \pmb{E}_\text{SI}\end{bmatrix}$. Therefore, $\pmb{H}_\text{SI}$ and $\pmb{G}_\text{sps}$ tend to be independent. Thus, the expected SI power of the spatial suppression (the worst case) is given by
	\begin{align}
	{\bar \Omega}_{\text{sps}(\epsilon_{_\text{SI}}\rightarrow\infty)}
	=\mathbb{E}\left\{||\pmb{w}_{\ku}^T\pmb{H}_\text{SI}\bar{\pmb{G}}_{\text{sps}}||^2\right\}
	=\mathbb{E}\left\{||\pmb{w}_{\ku}||^2\right\}.\label{sps_SI_w}
	\end{align}
	From (\ref{sps_SI_b}) and (\ref{sps_SI_w}), the approximation of  the expected SI power of spatial suppression is expressed as
	$
	{\bar \Omega}_\text{sps}\approx\frac{\mathbb{E}\left\{||\pmb{w}_{\ku}||^2\right\}}{1/\epsilon_{_\text{SI}}^2+1}.
	$
	This approximation approach is a simple application of harmonic mean.
	The approximation of the uplink SINR with spatial suppression is then given~by
	\begin{align}
	&\bar{\gamma}_{\text{sps}, \ku}^\text{UL}\nonumber
	\\&=\frac{(K_\text{UL}\rhoUL+1)(N-K_\text{UL}+1)}{K_\text{UL}(\sum_{\ell=1}^{K_\text{UL}}\frac{\rhoULell}{K_\text{UL}\rhoULell+1}+1)+\frac{\rho^{\text{SI}}}{\alpha_\text{anc}}\left(\frac{1}{1/\epsilon_{_\text{SI}}^2+1}\right)K_\text{UL}(1+\alpha_\text{tx})}.\label{SINR_sps}
	\end{align}

	\subsection{Expected SI Power in Correlated Channels}	\label{correl}

	In this subsection, we compare the expected SI power of the two SIC methods by considering the channel correlation.
	The estimated channel $\hat{\pmb{H}}$ is correlated when the channel is correlated while the estimation errors, elements of $\pmb{E}$, are uncorrelated~\cite{Jakob_TWC13}. 
	We first investigate how the correlated $\hat{\pmb{H}}_\text{SI}$ affects the level of SI power when $\pmb{G}_\text{sps}$ is used. 
	
	\begin{lemma}\label{lemma:corrSIch}
		When considering the correlated $\pmb{H}_\text{SI}$, $\zeta_\text{sps}$ is given~by 
		\begin{align}
		\zeta_\text{sps}=\text{Cov}(|E_{ik}|^2,|g_{km,\text{sps}}|^2)+\!\!\!\!\!\sum_{\ell=1,\ell\neq k}^{M}\!\!\!\!\!\epsilon_{_\text{SI}}^2r_{im,\text{sps}}\frac{\sqrt{qr_{k\ell,\text{SI}}^2+1}}{MK_\text{DL}},\nonumber
		\end{align}
		where $r_{im,\text{sps}}=\text{Corr}\left(E_{ik}E_{i\ell}^*, g_{km,\text{sps}}g_{\ell m,\text{sps}}^*\right)$ and $r_{k\ell,\text{SI}}=\text{Corr}\left(g_{km,\text{sps}}, g_{\ell m,\text{sps}}^*\right)$. In addition, $q$ is a constant that depends on the distribution of $|g_{km,\text{sps}}|^2$ (if the distribution is approximated to an exponential distribution, then $q=2$).
		\begin{IEEEproof}
			See Appendix C.
		\end{IEEEproof}
	\end{lemma}

	Because $r_{k\ell,\text{SI}}^2$ is proportional to the spatial correlation of the highly correlated SI channel, from \emph{Lemma}~\ref{proposition:zetaineq} and \emph{Lemma}~\ref{lemma:corrSIch}, we can conclude that as the correlation of the SI channel increases, the expected SI power of the spatial suppression decreases. Meanwhile, the correlation of the SI channel does not affect the expected SI power of the SI subtraction because $r_{im,\text{sps}}=0$.
	
	Next, we will show that the uplink performance of spatial suppression is better than that of the SI subtraction in the correlated channels with the practical angle distribution that is based on a Bessel function (e.g., Laplace distribution and wrapped Gaussian distribution~\cite{3GPP_SCM}).
	
	\begin{lemma}\label{lemma:correlatedchannel}
		In the correlated channels of the $\pmb{H}_\text{SI}$ and $\pmb{H}_\text{DL/UL}$, $\bar{\Omega}_\text{stt}\ge\bar{\Omega}_\text{sps}$ holds, which can be expressed as
		\begin{align}
		\bar{\Omega}_\text{sps}-\bar{\Omega}_\text{stt}&=\sum_{i=1}^N\mathbb{E}\left\{|w_{\ku i}|^2\right\}\sum_{m=1}^L\sum_{k=1}^M\zeta_{\text{sps}}\nonumber
		\\&~~+c\sum_{i=1}^N\sum_{j=1,j\neq i}^N \text{Cov}\left( w_{\ku i},w_{\ku j}^*\right)\le 0\nonumber,
		\end{align}
		where $c=\sum_{m=1}^L\sum_{k=1}^M\sum_{\ell=1}^M\mathbb{E}\left\{E_{ik}E_{j\ell}^*g_{km,\text{sps}}g_{\ell m,\text{sps}}^*\right\}$.
		\begin{IEEEproof}
			See Appendix D.
		\end{IEEEproof}
	\end{lemma}

\begin{figure*}[t!]
	\centering{\includegraphics[width=0.85\textwidth,keepaspectratio]{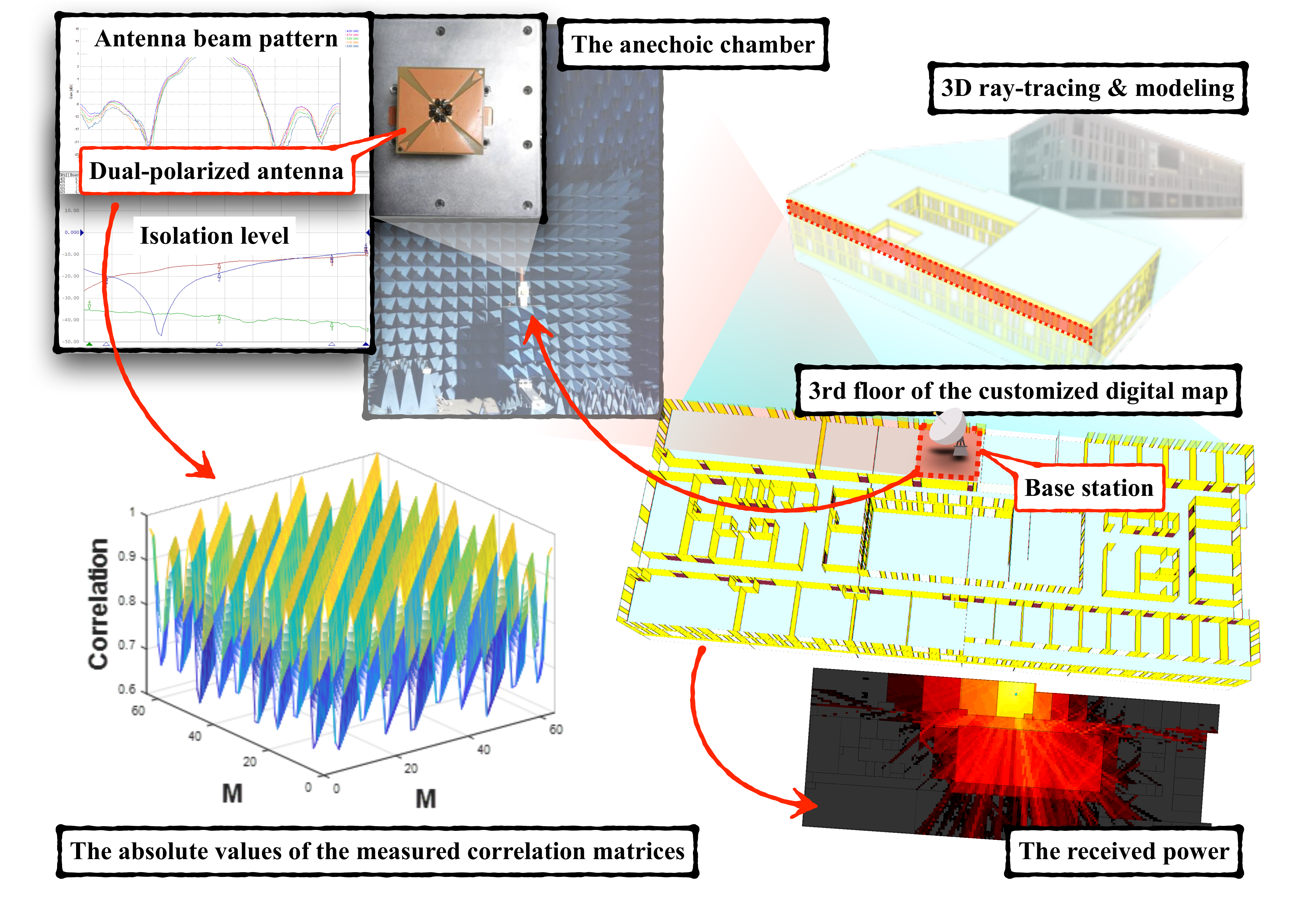}}
	\caption{The digital map for the 3D ray-tracing simulation and the absolute values of the measured $\pmb{R}_\text{TX(SI)}$ through the 3D ray-tracing simulation.}
	\label{Figure:digitalmap}
\end{figure*}

	\section{Performance Comparison with System Constraints}
	\label{sec:constraints}
	In this section, we compare the performance of the two SIC methods while considering the following system constraints: the downlink-to-uplink data traffic ratio under a non-full buffer scenario, the number of transmit and receive RF chains ratio with transmit power at the BS, and power scaling law at the BS. We hope that this discussion elucidates FD massive MIMO performance under various system constraints.
	
	\subsection{Optimal Antenna Ratio in a Non-Full Buffer Scenario}
	We first show that an asymmetric RF configuration is advantageous in a non-full-buffer asymmetric data traffic scenario. The traffic load density at the downlink and uplink are denoted by $\mu^\text{DL}$ and $\mu^\text{UL}$, and $T^\text{DL}=\mu^{DL}/\mathcal{R}^\text{DL}$; the transmission time for the given traffic loads is $T^\text{UL}=\mu^{UL}/\mathcal{R}^\text{UL}$, represented in ~\cite{Hong_Globecom17}. 
	We define the required downlink and uplink rates as the data rates for achieving $T^\text{DL}=T^\text{UL}=T$ in the FD mode within a certain time ($T$) to avoid half-duplex transmission, and their ratio is represented as $\eta$, which is the ratio of downlink traffic load to the uplink traffic load, $\eta=\mu^{DL}/\mu^{UL}$. Simple performance analysis for the optimal RF chain ratio is given as follows.
	
	\begin{remark}\label{remark:opt_ant}
		Assuming that the per-user data traffic is identical under asymmetric data traffic scenarios, the required downlink rate to uplink rate ratio can be expressed as $\eta=K_\text{DL}/K_\text{UL}$. In this scenario, we can obtain insight into the optimal ratio of the numbers of transmit RF chains and receive RF chains, and the optimal ratio of the SI subtraction and the spatial suppression given by $M/N=\eta$ and $(M-N)/N=\eta$, respectively, derived from the result in \cite{Min_TVT16}.
	\end{remark}

	\emph{Remark}~\ref{remark:opt_ant} implies that an asymmetric RF configuration is better than a symmetric RF configuration. In addition, this implication holds even if different per-user data traffic is assumed. It should be noted that the performance with the optimal RF chain ratio is not investigated in the following subsections, because the traffic ratio changes while the RF chain ratio at the BS is fixed. Instead, we present insights into the trade-off between the two SIC methods by considering arbitrary data traffic.
	
	\subsection{Performance Gap between Downlink and Uplink Rate with the Same Transmit Power}
	We assume that the numbers of downlink and uplink users are equal (i.e., $K=K_\text{DL}=K_\text{UL}$), while the downlink data traffic and uplink data traffic differ. In addition, the propagation gains have the same values for all $\kd$ and $\ku$. We analyze the performance from a per-user asymmetric data traffic perspective and	consider, for both links, high SINR regimes. SINRs can be expressed in terms of the received downlink SNR ($\varphi$, proportional to $\rho_\text{d}$), where the desired signal power is relatively higher than others at the downlink, and the SI power is relatively smaller than the desired power at the uplink. 
	For example, from (\ref{RDL_perfect1})-(\ref{RUL_perfect2}), $\varphi$ can be denoted by a linear function of $\rho_\text{d}$ ( $\varphi=\kappa\rho_\text{d}$) in the perfect channel estimation condition, where $\kappa=\frac{\betaDL}{K_\text{DL}(\sum_{\ku=1}^{K_\text{UL}}\rhoUU+1)}$. In these regimes, we can approximate the SINRs as follows: $\bar{\gamma}_{\text{stt}}^\text{DL}(\varphi)=(M-K)\varphi$, $\bar{\gamma}_{\text{sps}}^\text{DL}(\varphi)=(M-N-K)\varphi$, $\bar{\gamma}_{\text{stt}}^\text{UL}(\varphi)=\frac{\mathcal{S}}{\mathcal{I}+\omega_\text{stt}\varphi}$, $\bar{\gamma}_{\text{sps}}^\text{UL}(\varphi)=\frac{S}{\mathcal{I}+(\omega_\text{stt}-\omega_{_\varDelta})\varphi}$, where $\mathcal{S}$, $\mathcal{I}$, $\omega_\text{stt}$, and $\omega_\text{sps} (=\omega_\text{stt}-\omega_{_\varDelta})$ are scaled terms of the desired power, interference power, SI power of the SI subtraction, and SI power of the spatial suppression, respectively, by~$\varphi$:
	\begin{align}
	\bar{\gamma}_{\text{sps}}^\text{DL}(\varphi)=\frac{M-K-N}{M-K}\bar{\gamma}_{\text{stt}}^\text{DL}(\varphi)=\bar{\gamma}_{\text{stt}}^\text{DL}(\varphi)-\delta^\text{DL}\bar{\gamma}_{\text{stt}}^\text{DL}(\varphi),\nonumber\\
	\bar{\gamma}_{\text{stt}}^\text{UL}(\varphi)=\frac{\mathcal{I}+(\omega_\text{stt}-\omega_\varDelta)\varphi}{\mathcal{I}+\omega_\text{stt}\varphi}\bar{\gamma}_{\text{sps}}^\text{UL}(\varphi)=\bar{\gamma}_{\text{sps}}^\text{UL}(\varphi)-\delta^\text{UL}\bar{\gamma}_{\text{sps}}^\text{UL}(\varphi),
	\end{align}
	where $\delta^\text{DL}=\frac{N}{M-K}>0$, and $\delta^\text{UL}=\frac{\omega_\varDelta\varphi}{\mathcal{I}+\omega_\text{stt}\varphi}\ge0$. Thus, the approximations of the per-user rates are given by
	\begin{align}
	\mathcal{R}_\text{sps}^\text{DL}(\varphi)
	=\mathcal{R}_\text{stt}^\text{DL}(\varphi)+\log_2\left(1-\delta^\text{DL}\frac{\bar{\gamma}_{\text{stt}}^\text{DL}(\varphi)}{1+\bar{\gamma}_{\text{stt}}^\text{DL}(\varphi)}\right),\nonumber\\
	\mathcal{R}_\text{stt}^\text{UL}(\varphi)=\mathcal{R}_\text{sps}^\text{UL}(\varphi)+\log_2\left(1-\delta^\text{UL}\frac{\bar{\gamma}_{\text{sps}}^\text{UL}(\varphi)}{1+\bar{\gamma}_{\text{sps}}^\text{UL}(\varphi)}\right).
	\end{align}
	Thus, the gaps in the approximated rates between the two SIC methods at each link are expressed~as 
	\begin{align}
	\varDelta^\text{DL}(\varphi)=-\log_2\left(1-\delta^\text{DL}\frac{\bar{\gamma}_{\text{stt}}^\text{DL}(\varphi)}{1+\bar{\gamma}_{\text{stt}}^\text{DL}(\varphi)}\right)\ge 0,\nonumber\\
	\varDelta^\text{UL}(\varphi)=-\log_2\left(1-\delta^\text{UL}\frac{\bar{\gamma}_{\text{sps}}^\text{UL}(\varphi)}{1+\bar{\gamma}_{\text{sps}}^\text{UL}(\varphi)}\right)\ge 0.
	\end{align}
	By considering that $\mathcal{R}_\varDelta(\varphi)=\mathcal{R}_\text{stt}^\text{DL}(\varphi)-\mathcal{R}_\text{sps}^\text{UL}(\varphi)$ is the cross point when the gaps between the downlink and uplink rates of each SIC method are the same, we obtain
	\begin{align}
	\mathcal{R}_\text{stt}^\text{DL}(\varphi)-\mathcal{R}_\text{stt}^\text{UL}(\varphi)
	=\mathcal{R}_\varDelta(\varphi)+\varDelta^\text{UL}(\varphi),\nonumber\\
	\mathcal{R}_\text{sps}^\text{DL}(\varphi)-\mathcal{R}_\text{sps}^\text{UL}(\varphi)
	=\mathcal{R}_\varDelta(\varphi)-\varDelta^\text{DL}(\varphi),
	\end{align}
	and we can obtain the inequality as $\frac{\mathcal{R}_\text{stt}^\text{DL}}{\mathcal{R}_\text{stt}^\text{UL}}\ge\frac{\mathcal{R}_\text{sps}^\text{DL}}{\mathcal{R}_\text{sps}^\text{UL}}.\label{ieq:sic}$

	Therefore, because the gap between the downlink and uplink rates of the SI subtraction is higher than that of the spatial suppression for the given transmit power, the SI subtraction is more desirable when the gap between the downlink and uplink data traffic is higher (i.e., $\eta_\text{stt}\ge\eta_\text{sps}$), and vice versa.

	\subsection{Power Scaling Law in Downlink}
	
	Next, we consider a more practical constraint for asymmetric data traffic scenarios. In practical FD systems, achieving the maximum downlink rate can result in a bottleneck for supporting more downlink traffic and reducing the total transmission time of data traffic in the buffer at both links. Here, the maximum downlink rate is defined as the achievable rate by applying the best modulation order and coding rate.
	We denote it as $\tilde{\mathcal{R}}_\text{max}^\text{DL}$. In addition, we define the minimum required SNRs to achieve the maximum downlink rates as follows:
	\begin{align}
	\tilde\varphi_\text{stt} = \underset{\varphi}{\arg\min}\left\{ \tilde{\mathcal{R}}_\text{max}^\text{DL}\le\log_2(1+\bar{\gamma}^\text{DL}_\text{stt}(\varphi))\right\},\nonumber\\
	\tilde\varphi_\text{sps} = \underset{\varphi}{\arg\min}\left\{ \tilde{\mathcal{R}}_\text{max}^\text{DL}\le\log_2(1+\bar{\gamma}^\text{DL}_\text{sps}(\varphi))\right\},
	\end{align}
	where $\tilde\varphi_\text{stt}\le\tilde\varphi_\text{sps}$. 
	From the power scaling law in the downlink given by $\tilde{\mathcal{R}}_\text{max}^\text{DL}=\mathcal{R}_{\text{stt}}^\text{DL}(\tilde\varphi_\text{stt})=\mathcal{R}_{\text{sps}}^\text{DL}(\tilde\varphi_\text{sps})$, we have
	$
	\tilde\varphi_\text{stt}=\frac{M-K-N}{M-K}\tilde\varphi_\text{sps}.
	$
	We compare the uplink SINR of the SI subtraction at $\tilde\varphi_\text{stt}$ with that of the spatial suppression at $\tilde\varphi_\text{sps}$ as follows:
	\begin{align}
	\frac{\bar{\gamma}_{\text{stt}}^\text{UL}(\tilde\varphi_\text{stt})}{\bar{\gamma}_{\text{sps}}^\text{UL}(\tilde\varphi_\text{sps})}
	&=\frac{\mathcal{I}+(\omega_\text{stt}-\omega_\varDelta)\tilde\varphi_\text{sps}}{\mathcal{I}+\omega_\text{stt}\tilde\varphi_\text{stt}}
	=\frac{\mathcal{I}+(\omega_\text{stt}-\omega_\varDelta)\tilde\varphi_\text{sps}}{\mathcal{I}+\omega_\text{stt}\frac{M-K-N}{M-K}\tilde\varphi_\text{sps}}\nonumber\\
	&=\frac{\mathcal{I}+(\omega_\text{stt}-\omega_\varDelta)\tilde\varphi_\text{sps}}{\mathcal{I}+(\omega_\text{stt}-\frac{N}{M-K}\omega_\text{stt})\tilde\varphi_\text{sps}}.
	\end{align}
	In the correlated SI channel, the cross point of the uplink performance can be expressed as
	\begin{align}
	\mathcal{R}^\text{UL}_\text{cross}
	&=\frac{\omega_\varDelta}{\frac{N}{M-K}\omega_\text{stt}}\nonumber
	\\&=\underbrace{\frac{M-K}{N}}_{\substack{\text{trade-off from} \\ \text{an RF configuration}}}\cdotp\underbrace{\sum^{K}_m\sum^M_k\sum^M_\ell\frac{\sqrt{qr_{k\ell,\text{SI}}^2+1}}{MK}r_{im,\text{sps}}}_{\text{trade-off from the channel estimation error}}\label{eq:tradeoff},
	\end{align}
	where $\mathcal{R}^\text{UL}_\text{cross}\ge1$ represents ${\bar{\gamma}_{\text{sps}}^\text{UL}(\tilde\varphi_\text{sps})\ge\bar{\gamma}_{\text{stt}}^\text{UL}(\tilde\varphi_\text{stt})}$. 
	
	According to the discussion in subsection \ref{ieq:sic}, an increase in the number of transmit RF chains and the use of SI subtraction is desirable to support massive downlink traffic when the same transmit power is assumed. However, if the transmit power is scaled down to conserve energy, SI subtraction with such an RF configuration under performs the spatial suppression.
\begin{figure}[t!]
	\centering{\includegraphics[width=0.35\textwidth,keepaspectratio]{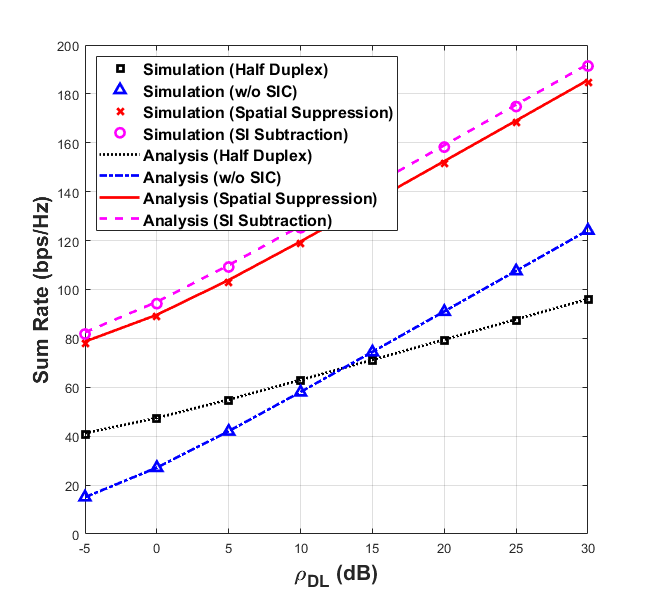}}
	\caption{Downlink/uplink sum rate with perfect channel estimation.}
	\label{Figure:perfect}
\end{figure}

\section{Performance Evaluation}\label{Section:C3_Section5}

\subsection{Simulation Parameters}
For performing numerical comparisons, we assumed that the FD BS has 88 RF chains ($M=64$, and $N=24$) and serves 12 users for each downlink and uplink.
The BS is equipped with dual-polarization antennas with a 15° tilt. We assume that 24 transmit RF chains and 24 receive RF chains are connected to horizontally polarized Port-1 and vertically polarized Port-2, respectively, while 40 transmit RF chains are connected to both ports. The uniform linear array with unit wavelength ($1 \lambda_\text{c}$) spacing is assumed at 2.52~GHz of the center frequency. We also assume that $\alpha_\text{anc}=30$~dB, which is reproduced from the cross-polarization ratio (XPR) of the dual-polarization antenna. We assume that the variance of the transmitter noise is large, and set it as $\alpha_\text{tx}=-10$~dB owing to the chip hardware cost for installing several RF chains. Finally, the simulation parameters are set as $\rho_{_\text{UL}}=10$~dB, $\beta^\text{SI}=-40$~dB, $\beta^\text{DL} = -80$~dB, $\beta^\text{UL} = -80$~dB, and $\beta^\text{UU} = -100$~dB, which are reproduced from the typical values of the path loss at the reference meter (1~m) and those in small cell environments.


\subsection{Channel Model for SI Channel}

\subsubsection{Scenario I: SI Channel with Simple Correlation Matrices}

The correlation matrices are simply modeled as
\begin{align}
\pmb{R}_\text{TX(SI)}=\begin{bmatrix}
1 & r_\text{TX,SI} & \dots & r_\text{TX,SI}^{M-2}& r_\text{TX,SI}^{M-1}\\
\vdots &  & \ddots &  & \vdots\\
r_\text{TX,SI}^{M-1} & r_\text{TX,SI}^{M-2} & \dots &  r_\text{TX,SI} & 1
\end{bmatrix},\nonumber
\end{align}
\begin{align}
\pmb{R}_\text{RX(SI)}=\begin{bmatrix}
1 & r_\text{RX,SI} & \dots & r_\text{RX,SI}^{N-2}& r_\text{RX,SI}^{N-1}\\
\vdots &  & \ddots &  & \vdots\\
r_\text{RX,SI}^{N-1} & r_\text{RX,SI}^{N-2} & \dots &  r_\text{RX,SI} & 1
\end{bmatrix}.\nonumber
\end{align}
According to \cite{3GPP_3DSCM}, we set $r_\text{TX/RX,SI}$ as 0.2 at a low correlation and 0.8 at a high correlation.

\begin{figure}[t!]
	\centering{\includegraphics[width=0.35\textwidth,keepaspectratio]{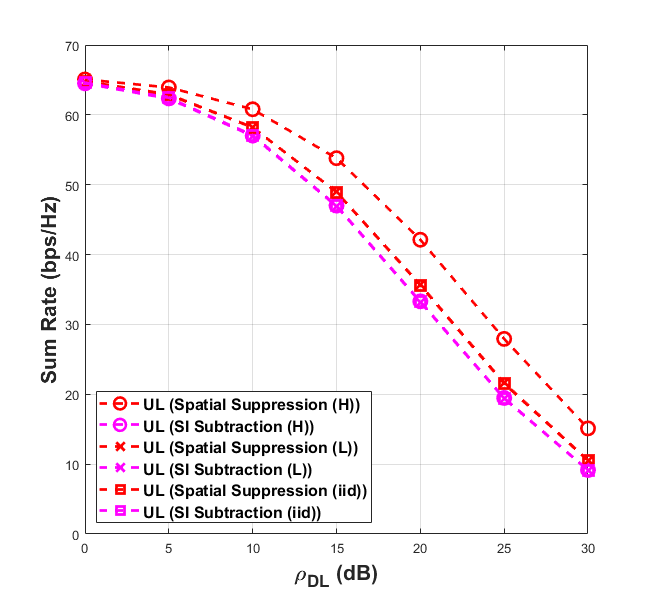}}
	\caption{Ergodic sum rate of the SI subtraction and the spatial suppression at high correlation and low correlation in the SI channel. The legends \emph{H} and \emph{L} denote high correlation and low correlation, respectively.}
	\label{Figure:RatevsSIcorrelation}
\end{figure}

\subsubsection{Scenario II: Practical SI Channel Measured from a 3D Ray-Tracing Tool}

\begin{figure*}[!h]
	\centering{\includegraphics[width=1\textwidth,keepaspectratio]{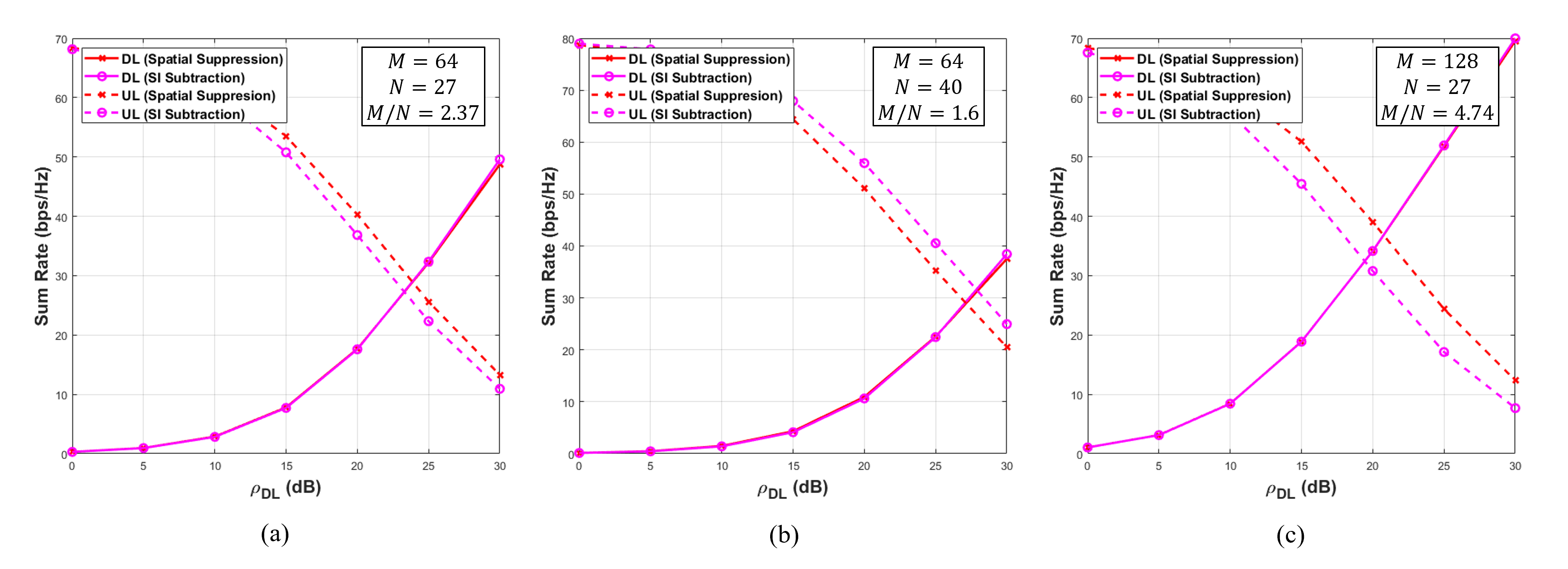}}
	\caption{Ergodic uplink sum rate of the SI subtraction and the spatial suppression at high correlation with different RF configuration: a) $M=64, N=27$; b) $M=64, N=40$; c) $M=128, N=27$.}
	\label{Figure:Tradeoff}
\end{figure*}

We develop a system-level simulator based on a 3D ray-tracing tool (WiSE, developed by Bell Labs~\cite{Wise}). We produce a customized digital map of a real building for an indoor scenario. Fig.~\ref{Figure:digitalmap} presents the test site from a third-floor perspective.
The digital map includes concrete walls and floors, metallic doors, glass windows, and sheetrock ceilings. The BS is located immediately under a 3~m-high
ceiling in a room facing the door to the south. We use both the co-polarization and cross-polarization patterns of the fabricated dual-polarization antenna, which are measured from an anechoic chamber~\cite{SM_ACCESS20}.\footnote{For modeling a more realistic dual-polarized channel, the XPR values of each ray are applied to co- and cross-polarization antenna patterns instead of using static XPR values in a polarization transfer matrix.} To realize an SI channel, we measure the power delay profiles of the $\ell$-th channel tap ($P_\ell$) and angle parameters of the $\ell$-th channel tap, such as the azimuth angle of arrival (AOA, $\phi_{\ell}^{\text{AOA}}$), azimuth angle of departure (AOD, $\phi_{\ell}^{\text{AOD}}$), zenith angle of arrival (ZOA, $\theta_{\ell}^{\text{ZOA}}$), and zenith angle of departure (ZOD, $\theta_{\ell}^{\text{ZOD}}$). The BS and users are equipped with the manufactured antenna and the isotropic antenna, respectively.


Next, we realized the SI channel in a realistic indoor scenario. We denote co-polarization and cross-polarization patterns of Port 1 by $G_{\text{rx}}^\text{V}$ and $G_{\text{rx}}^\text{H}$, respectively, and co-polarization and cross-polarization patterns of Port 2 by $G_{\text{tx}}^\text{H}$ and $G_{\text{tx}}^\text{V}$, respectively. Using a 3D ray-tracing simulation, we measured the $L_\text{tap}$ of the channel taps.
Based on a clustered channel model~\cite{3GPP_3DSCM}, the $(n,m)$-th element of the measured SI channel can be obtained as follows:
\begin{align}
[\pmb{H}_\text{SI}]_{nm}
&=\sum_{\ell=1}^{L_\text{tap}}P_\ell \begin{bmatrix}
G_{\text{rx}}^\text{H}\left(\theta_{\ell}^{\text{ZOA}},\phi_{\ell}^{\text{AOA}}\right) \\ G_{\text{rx}}^\text{V}\left(\theta_{\ell}^{\text{ZOA}},\phi_{\ell}^{\text{AOA}}\right)
\end{bmatrix}^T\nonumber\\
&\times
\begin{bmatrix}
\exp\left(j\Phi^\text{HH}_\ell\right)&\exp\left(j\Phi^\text{VH}_\ell\right)\\
\exp\left(j\Phi^\text{HV}_\ell\right)&\exp\left(j\Phi^\text{VV}_\ell\right)
\end{bmatrix}
\begin{bmatrix}
G_{\text{tx}}^\text{H}\left(\theta_{\ell}^{\text{ZOD}},\phi_{\ell}^{\text{AOD}}\right) \\ G_{\text{tx}}^\text{V}\left(\theta_{\ell}^{\text{ZOD}},\phi_{\ell}^{\text{AOD}}\right)
\end{bmatrix}\nonumber\\
&\times
\exp{\left(j\frac{2\pi}{\lambda_\text{c}}\pmb{\hat r}_{\text{rx},\ell}\pmb{\bar d}_{\text{rx},n}\right)}\exp{\left(j\frac{2\pi}{\lambda_\text{c}}\pmb{\hat r}_{\text{tx},\ell}\pmb{\bar d}_{\text{tx},m}\right)},\label{measuredSIchannel}
\end{align} 
where $\pmb{\bar d}_{\text{tx},m}$ and $\pmb{\bar d}_{\text{rx},m}$ are the location vectors of the transmit antenna $m$ and receive antenna $n$, respectively. In addition, $\pmb{\hat r}_{\text{tx},\ell}$ and $\pmb{\hat r}_{\text{rx},\ell}$ are spherical unit vectors with $\phi_{\ell}^{\text{AOA}}$ and $\theta_{\ell}^{\text{ZOA}}$ and with $\phi_{\ell}^{\text{AOD}}$ and $\theta_{\ell}^{\text{ZOD}}$, respectively. From (\ref{measuredSIchannel}), the measured correlation matrices can be calculated from $\pmb{R}_\text{TX(SI)}=\mathbb{E}[\pmb{H}_\text{SI}^*\pmb{H}_\text{SI}]$ and  $\pmb{R}_\text{RX(SI)}=\mathbb{E}[\pmb{H}_\text{SI}\pmb{H}_\text{SI}^*]$. The absolute values of the measured $\pmb{R}_\text{TX(SI)}$ are presented in Fig.~\ref{Figure:digitalmap}. We can confirm that the practical SI channel is highly correlated owing to the high direct path power and small angular spread.

\subsection{Numerical Result}


We first compare the proposed analysis and simulation results where perfect channel estimation is assumed at the BS. 
Fig.~\ref{Figure:perfect} shows that the results from~\ref{perfect} are approximately the same as the ergodic downlink/uplink sum rate of the FD-BS without SIC and with SI subtraction/spatial suppression. Because the SI is perfectly canceled out owing to perfect knowledge of the SI channel, both the uplink performance of the spatial suppression and the SI subtraction are identical, and they use the same ZF receiver. That is to say, there is a performance gap only in the downlink case.
An additional spatial domain $(N)$ is needed to cancel out the SI channel at the FD-BS with spatial suppression. Thus, we confirm that the SI subtraction outperforms the spatial suppression in the perfect channel estimation condition. For all the results below, imperfect channel estimation is assumed and simulated by using MATLAB.

\begin{figure}[t!] 
	\centering{\includegraphics[width=0.35\textwidth,keepaspectratio]{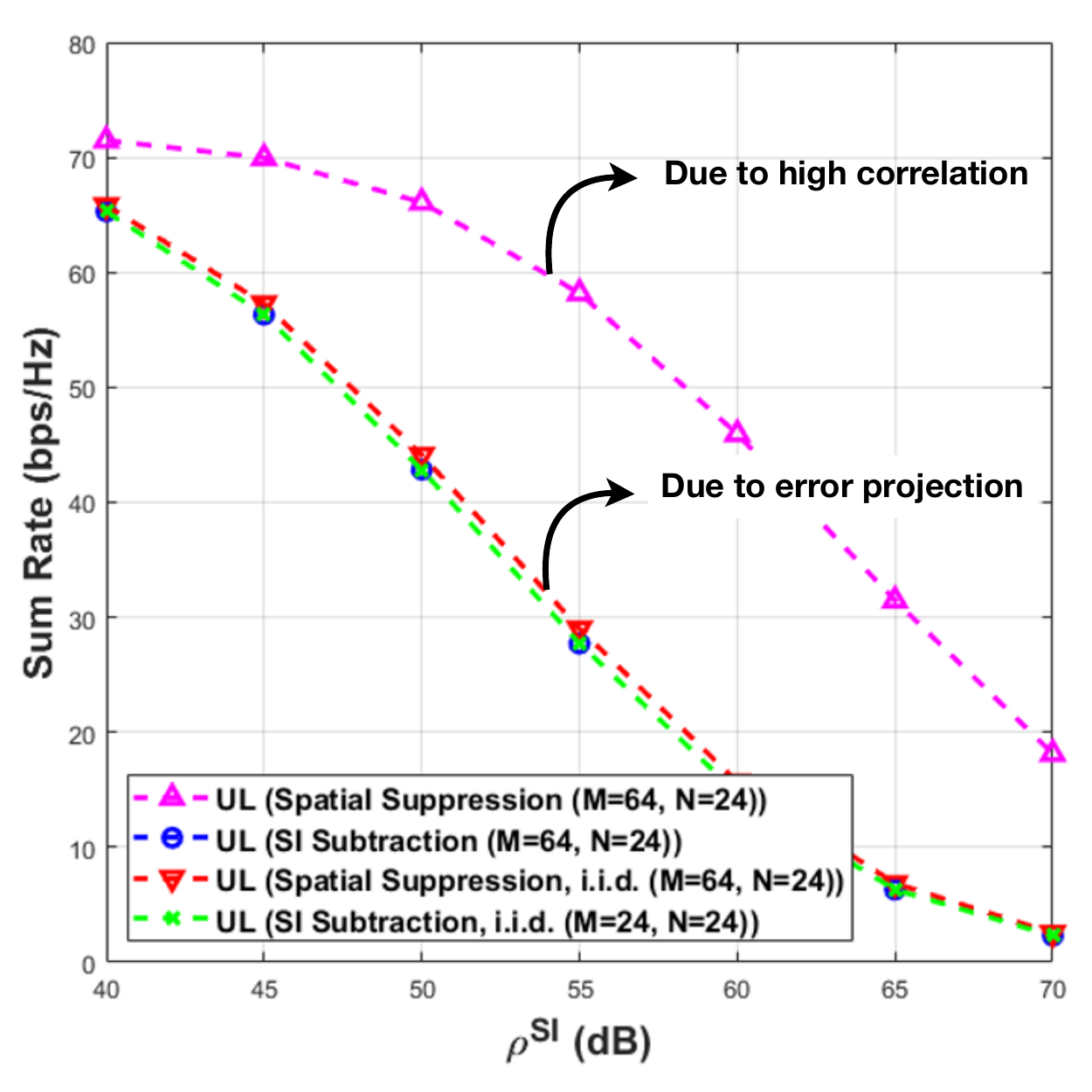}}
	\caption{Ergodic uplink sum rate of the SI subtraction and the spatial suppression in the practical SI channel.}
	\label{Figure:LLS1}
\end{figure}

\subsubsection{Scenario I}


Fig.~\ref{Figure:RatevsSIcorrelation} shows the ergodic uplink sum rate of the SI subtraction and spatial suppression at high and low correlations. The SI subtractions always exhibit the same performance irrespective of the correlation of the SI channel. Meanwhile, as the correlation of the SI channel increases, the performance of the spatial suppression increases owing to the projection onto the estimated SI channel manifold. This result verifies \emph{Lemma}~\ref{lemma:corrSIch}.

Fig.~\ref{Figure:Tradeoff} presents the ergodic uplink sum rate of the SI subtraction and the spatial suppression at high correlation with different RF configurations when the power scaling law at the downlink is assumed. A performance trade-off, as shown in (\ref{eq:tradeoff}), occurs in the RF configuration. In the uplink, spatial suppression outperforms the SI subtraction when $M/N$ is high, and vice versa. Under these constraints, both methods have the same downlink sum rate under these constraints. Furthermore, the opposite result is observed as compared to \emph{Remark}~\ref{remark:opt_ant}.

\begin{figure}[t!]
	\centering{\includegraphics[width=0.35\textwidth,keepaspectratio]{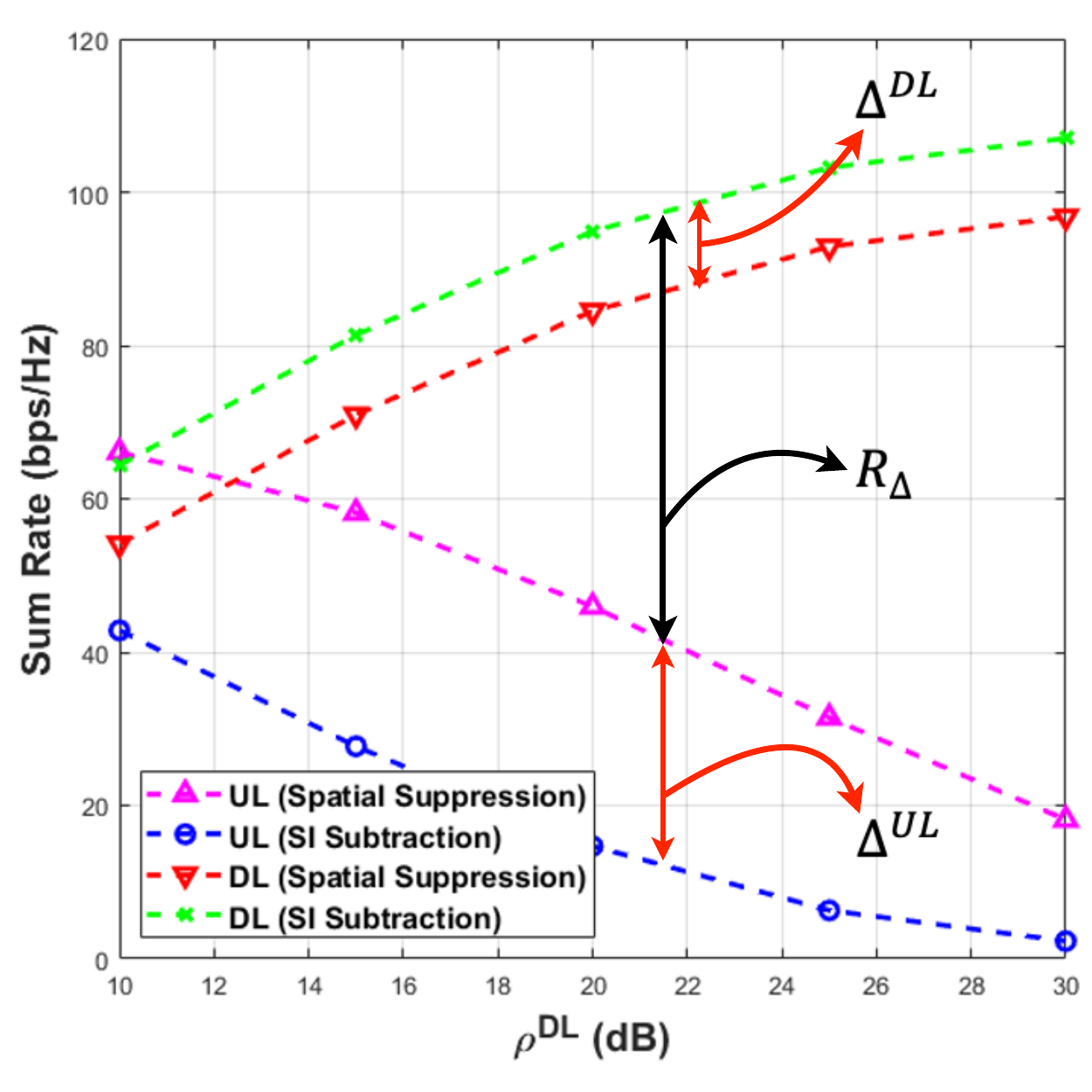}}
	\caption{The performance gap between the ergodic downlink sum rate and the ergodic uplink sum rate of the SI subtraction and the spatial suppression in the practical SI channel.}
	\label{Figure:LLS2}
\end{figure}

\subsubsection{Scenario II}


Fig.~\ref{Figure:LLS1} illustrates the ergodic uplink sum rate of the SI subtraction and spatial suppression when the measured SI channel is considered. To investigate the correlation impact, we additionally evaluated the sum rate with an i.i.d. SI channel. When $M$ increases and the correlation of the SI channel increases, no performance gain can be obtained by the SI subtraction. In addition, there is a significant performance gain with spatial suppression in the practical SI channel because of the high correlation.


Fig.~\ref{Figure:LLS2} presents the performance gap between the ergodic downlink sum rate and the ergodic uplink sum rate of the two SIC schemes. From the analysis in Section~\ref{sec:constraints}, $\varDelta^\text{DL}$ increases as $\delta^\text{DL}$ increases, which means that an additional spatial domain ($N$) is needed to cancel out the SI channel with spatial suppression. In addition, $\varDelta^\text{UL}$ increases as $\delta^\text{UL}$ increases, where $\delta^\text{UL}$ is proportional to $\epsilon_{_\text{SI}}$ and the correlation of the SI channel. Thus, the gap between the downlink and uplink rates of the SI subtraction is higher than that of the spatial suppression in the high SINR regimes in both links.

From the analysis and numerical results, we have shown which SIC method has better downlink and uplink sum rates in different channel conditions. In terms of total sum rates, Table~\ref{Tab.Const} summarizes the method most suitable under several given system constraints. 
	
	\begin{table}
	\centering
	\caption{The desired SIC technique under given system constraints with the corresponding formulas linked as references.\tablefootnote{It is worth noting that power scaling law in the downlink is assumed under the constraints marked with (*). Also (**) works when the gap between the downlink and uplink data traffic is greater (i.e., $\eta_\text{stt}\ge\eta_\text{sps}$).}}
	\begin{tabular}{c|c} \hline \hline
		\textbf{Given System Constraints}        & \textbf{Desired SIC Technique}         \\ \hline \hline
		Perfect Channel Estimation~(\ref{rate_stt}, \ref{rate_sps})  &  SI Subtraction
		\\	\hline
		Highly Correlated Channel (*)~(\ref{eq:tradeoff})  &  Spatial Suppression
		\\	\hline
		High $M/N$ (*)~(\ref{eq:tradeoff})  &  Spatial Suppression
		\\	\hline
		Total Transmit Power (**)~(\ref{ieq:sic})  &  SI Subtraction
		\\	\hline
	\end{tabular}
	\label{Tab.Const}
	\end{table}

		
		\section{Conclusions}
			\label{sec:conclusion}
		In this paper, two SIC methods in FD massive MIMO systems, i.e., SI subtraction and spatial suppression, have been investigated by considering the SI channel's estimation error and spatial correlation. In addition, we conducted an SI channel measurement campaign using a 3D ray-tracing simulation. The analysis and numerical results reveal that
		, in the uplink with imperfect channel estimation, the ergodic performance of the spatial suppression is better than that of the SI subtraction. This performance gain is caused by the correlation between the spatial suppression precoder and the estimation error of the SI channel. Moreover, we provided insight into which SIC algorithm is better under given system constraints, such as the downlink/uplink traffic ratio, number of RF chains, total transmit power, and power scaling law at the BS. 

%
	\appendix
	\subsection{Proof of Lemma 2}\label{Appendix Perfect}
	From \emph{Lemma}~\ref{decomposition}, all residual expectation terms in (\ref{eq:decompSI}) are zero because of the independent conditions that result in $\mathbb{E}\left\{ w_{\ku i}w_{\ku j}^*\right\}=\mathbb{E}\left\{ w_{\ku i}\right\}\mathbb{E}\left\{ w_{\ku j}^*\right\}=0$, $\text{Cov}(|a_{ik}|^2,|g_{km}|^2)=0$, and $\mathbb{E}\left\{a_{ik}a_{i\ell}^*g_{km}g_{\ell m}^*\right\}=0$, where $\ell\neq k$. Then, (\ref{eq:decompSI}) is given by
	\begin{align}
		\mathbb{E}\left\{||\pmb{w}_{\ku}^T\pmb{A}\pmb{G}||^2\right\}
		&=\sum_{i=1}^N\mathbb{E}\left\{ |w_{\ku i}|^2\right\}\sum_{m=1}^L\sum_{k=1}^M\mathbb{E}\left\{|a_{ik}|^2\right\}\mathbb{E}\left\{|g_{km}|^2\right\}\nonumber\\
		&=\sum_{i=1}^N\mathbb{E}\left\{ |w_{\ku i}|^2\right\}\sum_{m=1}^L\sum_{k=1}^M\sigma_A^2\mathbb{E}\left\{|g_{km}|^2\right\}\nonumber\\
		&\overset{(b)}{=}\sigma_A^2\mathbb{E}\left\{||\pmb{w}_{\ku}||^2\right\}.
	\end{align}
	Equation (\emph{b}) is derived based on $\sum_{i=1}^N\mathbb{E}\left\{ |w_{\ku i}|^2\right\}=\mathbb{E}\left\{||\pmb{w}_{\ku}||^2\right\}$ and $\sum_{m=1}^L\sum_{k=1}^M\mathbb{E}\left\{|g_{km}|^2\right\}=1$, owing to the definitions of $L^2$~norm and vector normalization.
	
	\subsection{Proof of Lemma 3}\label{proof:theorem:corrSIch}
	From $\mathbb{E}\left\{E_{ik}E_{i\ell}^*\right\}=0$ and $\mathbb{E}\left\{g_{km,\text{sps}}g_{\ell m,\text{sps}}^*\right\}=0$, $\zeta_{\text{sps}}$ is given by
	\begin{align}
		&\zeta_{\text{sps}}
		=\text{Cov}(|E_{ik}|^2,|g_{km,\text{sps}}|^2)\nonumber
		\\&~~+\sum_{\ell=1,\ell\neq k}^{M}\nonumber
		\\&\left[\text{Cov}(E_{ik}E_{i\ell}^*,g_{km,\text{sps}}g_{\ell m,\text{sps}})
		+\mathbb{E}\left\{E_{ik}E_{i\ell}^*\right\}\mathbb{E}\left\{g_{km,\text{sps}}g_{\ell m,\text{sps}}^*\right\}\right]\nonumber\\
		&=\text{Cov}(|E_{ik}|^2,|g_{km,\text{sps}}|^2)
		+\sum_{\ell=1,\ell\neq k}^{M}\text{Cov}(E_{ik}E_{i\ell}^*,g_{km,\text{sps}}g_{\ell m,\text{sps}}^*)\label{eq:zetacov}.
	\end{align}
	$\pmb{\bar G}_\text{sps}$ is the Pseudo-inverse of $\begin{bmatrix}\pmb{\hat H}_\text{DL}\\ \pmb{H}_\text{SI}+\pmb{E}_\text{SI}\end{bmatrix}$, the random variables, $E_{ik}$ and $g_{km,\text{sps}}$, tend to show opposite behaviors. Thus, each covariance term is negative as follows:
	\begin{align}
		\text{Cov}(|E_{ik}|^2,|g_{km,\text{sps}}|^2)\le 0, \nonumber 
	\end{align}
	\begin{align}
		\text{Cov}(E_{ik}E_{i\ell}^*,g_{km,\text{sps}}g_{\ell m,\text{sps}}^*)\le 0.\label{eq:cov2}
	\end{align}
	From (\ref{eq:zetacov})-(\ref{eq:cov2}), we obtain $\zeta_{\text{sps}}\le 0$.
	
	\subsection{Proof of Lemma 4}\label{proof:lemma4}
	Let $z_{k}$ and $z_{\ell}$ be uncorrelated random variables with the same distribution of $g_{km,\text{sps}}$. From the Cholesky decomposition, two correlated $g_{km,\text{sps}}$ and $g_{\ell m,\text{sps}}$ can be expressed as
	\begin{align}
		g_{km,\text{sps}}=z_{k},~
		g_{\ell m,\text{sps}}^*=r_{k\ell}z_{k}+\sqrt{1-r_{k\ell}^2}z_{\ell}.
	\end{align}
	Then, we obtain the variance of their multiplication as follows: 
	\begin{align}
		&\text{Var}\left\{g_{km,\text{sps}} g_{\ell\nonumber m,\text{sps}}^*\right\}\\&=\mathbb{E}\{|g_{km,\text{sps}}|^2|g_{\ell m,\text{sps}}|^2\}+[\mathbb{E}\{g_{km,\text{sps}}g_{\ell m,\text{sps}}^*\}]^2\nonumber\\
		&=r_{k\ell}^2\mathbb{E}\{z_{k}\}+2r_{k\ell}\sqrt{1-r_{k\ell}^2}\mathbb{E}\{|z_k|^2z_kz_\ell\}+r_{k\ell}\mathbb{E}\{|z_k|^2\}\nonumber\\
		&~~+(1-r_{k\ell}^2)\mathbb{E}\{|z_k|^2|z_\ell|^2\}+\sqrt{1-r_{k\ell}^2}\mathbb{E}\{z_kz_\ell\}\nonumber\\
		&=r_{k\ell}^2\mathbb{E}\{|g_{km,\text{sps}}|^4\}+\mathbb{E}\{|g_{km,\text{sps}}|^2\}^2,
	\end{align}
	where $\mathbb{E}\{|z_k|^2z_kz_\ell\}=0$, and $\mathbb{E}\{z_kz_\ell\}=0$.
	Thus, when $g_{km,\text{sps}}$ and $g_{\ell m,\text{sps}}$ are correlated, the variance of their multiplication is given~by
	\begin{align}
		\text{Var}\left\{g_{km,\text{sps}} g_{\ell m,\text{sps}}^*\right\}=r_{k\ell}^2\mathbb{E}\{|g_{km,\text{sps}}|^4\}+\mathbb{E}\{|g_{km,\text{sps}}|^2\}^2.\label{eq:varcorrsps}
	\end{align}
	From (\ref{eq:zetacov}) and (\ref{eq:varcorrsps}), we obtain
	\begin{align}
		\zeta_{\text{sps}}
		&=\text{Cov}(|E_{ik}|^2,|g_{km,\text{sps}}|^2)~~\nonumber
		\\&~~+\!\!\!\!\!\sum_{\ell=1,\ell\neq k}^{M}\!\!\!\!\!\text{Cov}\left\{E_{ik}E_{i\ell}^*,g_{km,\text{sps}}g_{\ell m,\text{sps}}^*\right\}\nonumber\\
		&=\text{Cov}(|E_{ik}|^2,|g_{km,\text{sps}}|^2)\nonumber
		\\&~~+\!\!\!\!\!\sum_{\ell=1,\ell\neq k}^{M}\!\!\!\!\!r_{im,\text{sps}}\sqrt{\text{Var}[E_{ik}E_{i\ell}^*]\text{Var}[g_{km,\text{sps}}g_{\ell m,\text{sps}}]}\nonumber\\
		&=\text{Cov}(|E_{ik}|^2,|g_{km,\text{sps}}|^2)\nonumber
		\\&~~+\sum_{\ell=1,\ell\neq k}^{M}\epsilon_{_\text{SI}}^2r_{im,\text{sps}}\sqrt{r_{k\ell}^2\mathbb{E}\{|g_{km,\text{sps}}|^4\}+\mathbb{E}\{|g_{km,\text{sps}}|^2\}^2}\nonumber\\
		&=\text{Cov}(|E_{ik}|^2,|g_{km,\text{sps}}|^2)+\!\!\!\!\!\sum_{\ell=1,\ell\neq k}^{M}\epsilon_{_\text{SI}}^2r_{im,\text{sps}}\frac{\sqrt{qr_{k\ell,\text{SI}}^2+1}}{MK_\text{DL}},
	\end{align}
	where $\mathbb{E}[|g_{km}|^2]=\frac{1}{MK}$ and $\mathbb{E}[|g_{km}|^4]=\frac{q}{(MK)^2}$.
	
	\subsection{Proof of Lemma 5}\label{proof:lemma5}
	We assume $b^{(+)}_k=\text{Sort}(|\text{Cov}\left( w_{\ku i},w_{\ku j}^*\right)|)$ for $\text{Cov}\left( w_{\ku i},w_{\ku j}^*\right)\ge 0$ and $b^{(-)}_\ell=\text{Sort}(|\text{Cov}\left( w_{\ku i},w_{\ku j}^*\right)|)$ for $\text{Cov}\left( w_{\ku i},w_{\ku j}^*\right)\le 0$. For practical angle distributions, the correlation function in terms of antenna spacing can be expressed by a zero-order Bessel function~\cite{Corr_Model}. The main lobe of a Bessel function is larger than the side lobes, so $b^{(+)}_1>b^{(-)}_1$ holds. 
	Thus, in the correlated $\pmb{H}_\text{UL}$, we can obtain
	\begin{align}
		\sum_{i=1}^N\sum_{j=1,j\neq i}^N \text{Cov}\left( w_{\ku i},w_{\ku j}^*\right)\ge 0.\label{cov}
	\end{align}

	From~\emph{Lemma}~\ref{decomposition}, we check the quadratic forms of $\bar{\Omega}_\text{stt}$ and $\bar{\Omega}_\text{sps}$ as follows:
	\begin{align}
		{\bar \Omega}_{\text{stt}}=\sum_{i=1}^N\mathbb{E}\left\{|w_{\ku i}|^2\right\}\sum_{m=1}^L\sum_{k=1}^M\mathbb{E}\left\{|E_{ik}|^2\right\}\mathbb{E}\left\{|g_{km, \text{ZF}}|^2\right\},\label{eq:quadstt}
	\end{align}
	where $\bar{\Omega}_{\pmb{E}_\text{SI},(i\neq j),\text{stt}}=0$, since $\mathbb{E}\left\{E_{ik}E_{j\ell}^*g_{km,\text{ZF}}g_{\ell m,\text{ZF}}^*\right\}=\mathbb{E}\left\{E_{ik}E_{j\ell}^*\right\}\mathbb{E}\left\{g_{km,\text{ZF}}g_{\ell m,\text{ZF}}^*\right\}=0$ when $i\neq j$,
	\begin{align}
		&{\bar \Omega}_{\text{sps}}\nonumber
		\\&=\sum_{i=1}^N\sum_{j=1}^N\mathbb{E}\left\{ w_{\ku i}w_{\ku j}^*\right\}\sum_{m=1}^L\sum_{k=1}^M\sum_{\ell=1}^M\mathbb{E}\left\{E_{ik}E_{j\ell}^*g_{km,\text{sps}}g_{\ell m,\text{sps}}^*\right\}.\label{eq:quadsps}
	\end{align}
	From (\ref{eq:quadstt}), the expected SI power term of the SI subtraction is not affected by the correlated~$\pmb{H}_\text{DL}$. 
	Similar to (\ref{eq:zetacov}), we have $\mathbb{E}\left\{E_{ik}E_{j\ell}^*g_{km,\text{sps}}g_{\ell m,\text{sps}}^*\right\}=\text{Cov}\left(E_{ik}E_{j\ell}^*,g_{km,\text{sps}}g_{\ell m,\text{sps}}^*\right)$. Then, we can obtain
	\begin{align}
		c=\sum_{m=1}^L\sum_{k=1}^M\sum_{\ell=1}^M\mathbb{E}\left\{E_{ik}E_{j\ell}^*g_{km,\text{sps}}g_{\ell m,\text{sps}}^*\right\}\le0,\label{c}
	\end{align} 
	where $i \neq j$.
	From (\ref{eq:quadstt}) and (\ref{eq:quadsps}), the gap between $\bar{\Omega}_\text{stt}$ and $\bar{\Omega}_\text{sps}$ in the correlated channels can be derived as
	\begin{align}
		\bar{\Omega}_\text{sps}-\bar{\Omega}_\text{stt}
		&=\sum_{i=1}^N\mathbb{E}\left\{|w_{\ku i}|^2\right\}\sum_{m=1}^L\sum_{k=1}^M\zeta_{\text{sps}}\nonumber\\
		&~~+\sum_{i=1}^N\sum_{j=1,j\neq i}^N\mathbb{E}\left\{ w_{\ku i}w_{\ku j}^*\right\}\nonumber\\
		&~~\times\sum_{m=1}^L\sum_{k=1}^M\sum_{\ell=1}^M\mathbb{E}\left\{E_{ik}E_{j\ell}^*g_{km,\text{sps}}g_{\ell m,\text{sps}}^*\right\}\nonumber\\
		&=\sum_{i=1}^N\mathbb{E}\left\{|w_{\ku i}|^2\right\}\sum_{m=1}^L\sum_{k=1}^M\zeta_{\text{sps}}\nonumber\\
		&~~+c\sum_{i=1}^N\sum_{j=1,j\neq i}^N \text{Cov}\left( w_{\ku i},w_{\ku j}^*\right).\label{eq:gapOmega}
	\end{align}
	From~(\ref{cov}),~(\ref{c}), and \emph{Lemma}~\ref{proposition:zetaineq}, (\ref{eq:gapOmega}) can be expressed as $\bar{\Omega}_\text{sps}-\bar{\Omega}_\text{stt}\le 0$.
	
%
%
	\bibliographystyle{IEEEtran}
	\bibliography{FD_reference_v2}
	
\end{document}